\documentclass[useAMS,usenatbib,fleqn]{mn2e}

\usepackage[T1]{fontenc}
\usepackage{ae,aecompl}

\usepackage{graphicx}
\usepackage{amsmath}
\usepackage{amssymb}

\usepackage[FIGTOPCAP]{subfigure}

\usepackage{upgreek}
\usepackage{bm}

\title[Spirals in
  Triaxial Dark Matter Haloes]{Stellar Spiral Structures in Triaxial Dark Matter Haloes}
\author[Hu \& Sijacki]{Shaoran Hu$^1$, Debora Sijacki$^1$\\
$^1$ Institute of Astronomy and Kavli Institute for Cosmology, University of
Cambridge, Madingley Road, Cambridge CB3 0HA, UK}

\begin{document}
\label{firstpage}
\pagerange{\pageref{firstpage}--\pageref{lastpage}}

\maketitle

\begin{abstract}
  We employ very high resolution simulations of isolated Milky Way-like
  galaxies to study the effect of triaxial dark matter haloes on exponential
  stellar discs. Non-adiabatic halo shape changes can trigger two-armed
  grand-design spiral structures which extend all the way to the edge of the
  disc. Their pattern speed coincides with the inner Lindblad resonance
  indicating that they are kinematic density waves which can persist up to
  several Gyr. In dynamically cold discs, grand-design spirals are swing
  amplified and after a few Gyr can lead to the formation of (multi-armed)
  transient recurrent spirals. Stellar discs misaligned to the principal
  planes of the host triaxial halo develop characteristic integral shaped
  warps, but otherwise exhibit very similar spiral structures as aligned discs.
  For the grand-design spirals in our simulations, their strength dependence
  with radius is determined by the torque on the disc, suggesting that by
  studying grand-design spirals without bars it may be possible to set
  constraints on the tidal field and host dark matter halo shape. 
\end{abstract}

  
\section{Introduction}
\renewcommand{\thefootnote}{\fnsymbol{footnote}}
\footnotetext[1]{E-mail: sh759@ast.cam.ac.uk}

For many years spiral structures in galaxies have been the subject of
extensive observational,  theoretical and numerical studies, but their origin
still remains unclear. While the morphologies of spiral structures vary
considerably, they can be generally classified into two broad categories:
`grand-design'  spirals with mostly two arms that extend over a large range
of radii and `flocculent' spirals that consist of many small fragments of
arms. Both kinds of spirals are mostly trailing rather than leading.

Theories of formation and evolution of spirals fall into two categories:
(a) self-induced spiral formation, in which spiral structures form due to
gravitational interaction between finite number of stars, and (b)
externally driven spiral formation, in which spiral structures form as a
response to external perturbations.

Lindblad proposed a theory of quasi-stationary spiral structures for
grand-design spirals \citep[e.g.][]{Lindblad1963}. In his theory, spiral
structures are in fact kinematic density waves with a pattern speed of
$\Omega_\mathrm{P}=\Omega-\kappa/n$, where $\Omega$ is the corotation
  velocity, $\kappa$
is the epicyclic frequency and $n$ is the number of arms (typically equal to 2).
In a frame moving
with this pattern speed, the orbits of the epicyclic motion of the stars
become ellipses without precession, which guarantees that the density field of
the disc remains constant in this rotating frame. When orbits are arranged in
a way that the highest density falls into a spiral-like shape
\citep[e.g.][]{kalnajs1973}, constantly rotating stationary grand-design
spiral structures can survive in the disc \citep{Lindblad1956}.
\citet{lin1964} developed a theory of density waves of quasi-stationary spiral
structures further.  They regarded spiral structures as large-scale waves
propagating through the disc in a linear regime \citep[see
  also][]{Bertin1996}.

 `Flocculent' spiral structures (and some of the `grand-design'
spirals) are considered to be caused by instabilities due to self-gravity. To
form spiral structures of this kind, two important processes are needed: swing
amplification of small perturbations and a feedback loop. The local stability
of a razor-thin disc with respect to axisymmetric tightly wound perturbations is characterized by
Toomre's $Q$ parameter \citep{toomre1964}.  For stellar discs, we have
\begin{equation}
  Q=\frac{\sigma_R\kappa}{3.36 \,G \Sigma},
\end{equation}
where $\sigma_R$ is the velocity dispersion in the radial direction and $\Sigma$ is the surface density. When $Q>1$,
the disc is locally stable to axisymmetric perturbations, while for $Q<1$, the
disc becomes locally unstable. Considering the case of non-axisymmetric
perturbations, \citet{Julian1966} studied the stability of differentially
rotating discs and showed that due to self-gravity, small perturbations can be
greatly amplified.  \citet{toomre1981} explored the theory further and showed that strong
swing amplification can occur when $Q$ is slightly higher than 1, i.e. if the
disc is locally stable but the self-gravity is still strong and the wavelength
ratio $X$ is appropriate (typically between 1 and 3):
\begin{equation}
X=\frac{k_\mathrm{crit}R}{m}=\frac{\kappa^2R}{2\uppi G\Sigma m}\,,\label{eq:3}
\end{equation}
where $R$ is the radius, $m$ is the number of arms and
$k_\mathrm{crit}=\kappa^2/(2\uppi G\Sigma)$ is the critical wavelength of the
swing amplification. In this  process small leading waves are  amplified to
strong trailing waves and the amplification factor can be as high as
100. \citet{toomre1981} also demonstrated that swing amplification can act on
global spiral patterns and greatly amplify their strength, though external
torques, for example, are sufficient to form those patterns. Similarly,
\citet{Grand2012A} showed that two-armed spiral structures in barred galaxies can be
dominated by swing amplification.

Recent studies combining numerical and analytic efforts have further shown 
that stars interact with each other over long periods of time through resonances. 
\citet{Sellwood2012} found that stars are scattered at the inner Lindblad 
resonance as the transient spiral structures form, leading to a 
re-distribution in the action space. When stars are rotated randomly to erase 
non-axisymmetric features without affecting the distribution in action space,
spiral structures restore rapidly, indicating that the scattering at inner 
Lindblad resonance is more fundamental than the change in the density field.
Such a scattering has been recently studied by \citet{Fouvry2015} using a
dressed Fokker--Planck formalism, which offers a powerful tool to probe the evolution  
of discs as a function of their properties.

While a number of past numerical studies found that spirals fade out quickly
over time \citep[for a recent review see][and references therein]{Dobbs2014},
recent studies showed that in a stellar disc with more than a few million
particles spiral arms persist for longer periods of time, indicating
that previous results were suffering from discreteness effects \citep[e.g.][]{Fujii2011,Grand2012}.
\citet{DOnghia2013} showed that with a sufficiently high number of particles
(of the order of $10^8$), stellar discs with $Q$ slightly larger than $1$ can
stay stable over a few galactic years. When, however, density perturbations
are introduced in the disc (in the form of heavy particles), a transient
spiral pattern forms which itself can act as the source of newly formed spiral
arms. \citet{sellwood2014} studied the power spectra of such transient spirals,
and found that they are superposition of several rigid rotating modes lying
between the inner and the outer Lindblad resonance. They also found that such
waves scatter particles towards new regions of a disc, thus changing the
impedance of the disc, reflecting the waves and hence giving rise to new
standing waves. In fact, \citet{fouvry2015d} showed that such
  simulations of spiral structures dominated by self-gravity in discrete discs can
 be characterized by the Balescu--Lenard equation, whose predictions on
 the properties of the secular orbital diffusion agree very well with simulations. 

Taken together, these recent numerical works
indicate that to study stellar spiral structures that may form in response to external
perturbations, discs need to be represented
with a very high number of resolution elements, to both minimize arteficial
spiral heating and the Poisson noise in the initial conditions,
  ensuring that 
the growing time of transient spirals is long enough compared to the evolution 
of grand-design spirals.

For the triggering mechanism of the grand-design spiral structures,
additionally to bars \citep[e.g.][]{Salo2010,Athanassoula2012} and close
companions \citep[e.g.][]{Purcell2011}, torques caused by the host dark
matter halo have been invoked as well. Even though the properties of dark
matter haloes, such as their exact shape and mass distribution (and the
back-reaction of baryons), are still not precisely known, already early
work \citep[e.g.][]{Binney1978, barnes1987, frenk1988} have indicated that the
haloes are generally triaxial.  Follow-up studies with a variety of
configurations, including dispersionless gravitational collapse models
\citep{warren1992}, self-interacting dark matter models \citep{yoshida2000},
haloes formed in different cosmological models \citep{jing1995, thomas1998},
and more recent higher-resolution studies \citep[e.g.][]{Bryan2013, Zhu2015}
all find that the dark matter haloes are triaxial. Several generalizations of
analytical, spherical halo models that include halo triaxiality were also
proposed \citetext{\citealp{jing2002}; \citealp{bowden2013}}. 
 Analysing high resolution Aquarius simulations \citep{Springel2008}, \citet{Vera-ciro2011} showed that due 
to the cosmic growth history of dark matter haloes, their triaxiality 
can change rapidly over time, implying that the disc will be subject to a time-dependent torque. It is generally
believed that the inclusion of baryons leads to a reduction of halo
triaxiality \citep{Dubinski1994,  DOnghia2010, zemp2012, Bryan2013, Zhu2015},
with haloes becoming more oblate, especially in the central region. However,
the resulting mildly triaxial halo mass distribution may still impart a
significant torque on to the disc of the central galaxy.

In fact, \citet{debuhr2012} have found that the gravitational potential of a
disc can flatten the halo, while in return bars and warps can develop in the
disc under the influence of the flattened halo. For grand-design spiral
structures, \citet{dubinski2009} studied the impact of external torque on the
disc. In their simulations, it is assumed that while the inner part of the dark
matter halo is aligned with the disc, the outer region is misaligned and
tumbling, causing an external torque. Under such external torque,
\citet{dubinski2009} found that grand-design two-armed spiral structures can
develop in the disc, along with warps and bars. More recently,
\citet{Khoperskov2013} found that grand-design spiral patterns can form in
discs within haloes which are gradually turned from spherical into triaxial
\citep[see also][]{Khoperskov2015}.   

None the less, the relation between the triaxiality of dark matter haloes and
the spiral structures in discs is still not fully understood.  Therefore, a
careful study employing very high resolution discs is needed to understand the
influence of halo shapes on the disc structure which is the aim of this
paper. Also since there are two different kinds of spiral structures,
grand-design and flocculent ones, it is important to understand how they form
in triaxial haloes.

The paper is organized as follows. In Section~\ref{sec:method}, we introduce
the methodology  together with galaxy and halo models we use. Our results are then presented in
Section~\ref{sec:results}. In Section~\ref{sec:finite-resol-effects}, we
examine the numerical effects caused by the finite resolution of simulations
which act as perturbations of the density field of stellar discs (see
  also Appendix~\ref{app:dssh}). In
Section~\ref{sec:time-depend-triax}, we study the effect of how triaxial haloes
are introduced into the system with very high resolution simulations, 
  while in Section~\ref{sec:triax-halo-cosm} we study spiral pattern generated
by time-dependent triaxial haloes as predicted by cosmological simulations. We
then  
focus on the impact of triaxial haloes of different shapes on the discs in
Section~\ref{sec:depend-spir-strength}. In Section~\ref{sec:swing-ampl-spir},
we discuss the nature of transient spirals that emerge out of grand-design
spiral structures due to non-linear effects. The underlying mechanism of the
grand-design spirals is then studied in Section~\ref{sec:formation-mechanism} (for discs
misaligned with the major axes of the halo, see
Appendix~\ref{sec:discs-misal-triax}).  Finally, in
Section~\ref{sec:conclusions} we summarize our results. 

\section{Method}
\label{sec:method}
\subsection{The Numerical Approach}
\begin{table}
  \centering
  \caption{Numerical parameters for simulations with different number of star
    particles. For simulations listed below, the total mass of the stars  is
    either $M_\mathrm{*}=1.9\times 10^{10}M_\odot$ (low-$Q$ discs) or
    $M_\mathrm{*}=9.5\times 10^{9}M_\odot$ (high-$Q$ discs). Number of star
    particles $N$, mass of a single star particle, $m$, and gravitational
    softening length, $\epsilon_\mathrm{grav}$, are listed below.}
  \begin{tabular}{rlll}
    \hline
$M_\mathrm{*}(M_\odot)$&$N$ & $m(M_\odot)$ & $\epsilon_\mathrm{grav}(\mathrm{pc})$\\
    \hline
$1.9\times 10^{10}$&$10^5$ & $1.9\times 10^5$ & $163$\\
$1.9\times 10^{10}$&$10^6$ & $1.9\times 10^4$ & $76$\\
$1.9\times 10^{10}$&$10^7$ & $1.9\times 10^3$ & $35$\\
$1.9\times 10^{10}$&$10^8$ & $1.9\times 10^2$ & $16$\\
$9.5\times 10^{9}$&$10^5$ & $9.5\times 10^4$ & $163$\\
$9.5\times 10^{9}$&$10^6$ & $9.5\times 10^3$ & $76$\\
$9.5\times 10^{9}$&$10^7$ & $9.5\times 10^2$ & $35$\\
$9.5\times 10^{9}$&$10^8$ & $95$ & $16$\\
\hline
  \end{tabular}
  \label{tab:2}
\end{table}

We perform simulations of stellar discs embedded in different dark matter halo
models with {\sc gadget-3}, whose previous version {\sc gadget-2} is
described in \citet{springel2005cosmological}. {\sc gadget-3} is an
$N$-body/smoothed particle hydrodynamics code. In the code, stars are represented by a finite number of
stellar particles. Our choice of the number of star particles varies from
$10^5$ to $10^8$, so for a Milky Way-like galaxy a single star particle in the
simulation typically represents about $10^6$--$10^3$ stars. Their dynamics is simulated with the $N$-body algorithm.

Gaseous component is very important in the evolution of galactic discs. By
interacting with the stars through gravity, gas can enhance the self-gravity
of the disc. Also it can develop sharp shocks when placed in a gravitational
potential caused by spiral structures \citep{gittins2004, Dobbs2006,
  Dobbs2007}. Moreover, new stars  formed out of the high-density gas clumps
can act as a cooling source to the dynamical temperature of the disc by
lowering the velocity dispersion. However, modelling gas numerically is very
difficult, given that complex cooling and heating processes need to be taken
into account. Therefore, we do not include gaseous component in this work.

Analytic representation of dark matter haloes is employed in all of the
simulations. Given that the total mass of the dark matter halo is much larger
than the total mass of the disc, if we represent dark matter halo with
particles, we need much more  dark matter halo particles than star particles
(a number that turns out to be prohibitively large!), so that the Poisson
noise in the halo is comparable to that in the disc. In fact, for testing
purposes, we ran a simulation of a live disc within a live halo, both of them
with $10^8$ particles. In this simulation, strong transient spiral structures
form almost instantly, making it impossible to study the possible effects of
the halo shape on the disc. To avoid this effect, we would need much more dark
matter particles.  However, we are primarily interested in the behaviour of
the disc rather than that of the dark matter halo. To make sure that the halo
does not induce numerical artefacts to the system and to direct computational
resources on the object we are interested in,  we used  analytic dark matter
halo models rather than the dark matter particles. It is worth mentioning that
with our methodology we cannot study the back-reaction of the disc on to the dark
matter halo. To somewhat mitigate this issue, we study static dark matter
haloes of different shapes.

For computation of the gravity, the {\sc gadget-3} code employs the TreePM
method \citep{springel2005cosmological}. The combination of the two methods,
the Tree method and the Particle-Mesh (PM) method, gives us high efficiency in
calculating gravitational forces with high accuracy.  Constant gravitational
softening length $\epsilon_\mathrm{grav}$ for star particles is used in all
simulations. Typical values of $\epsilon_\mathrm{grav}$ used in our
simulations are shown in Table~\ref{tab:2}.

\subsection{Modelling of Stellar Discs}

We set up our disc model following the description in
\citet{springel2005}. The disc has an exponential surface density profile and
an isothermal sheet profile vertically, described by  
\begin{equation}
  \label{eq:dm}
  \rho_\mathrm{*}(R,z)=\frac{M_\mathrm{*}}{4\uppi z_0 R_\mathrm{S}^2} \mathrm{sech}^2\left(\frac{z}{z_0}\right)\exp\left(-\frac{R}{R_\mathrm{S}}\right),
\end{equation}
where $R_\mathrm{S}=3.13\,\mathrm{kpc}$ is the scalelength of the disc,
$M_\mathrm{*}$ is the total mass of the stars  and $z_0=0.1 R_\mathrm{S}$ is
the scaleheight. The total mass of the system,
$M_\mathrm{T}=M_*+M_\mathrm{dm}$ is $9.5\times 10^{11}M_\odot$ in all our
simulations. However, two kinds of discs with different disc mass ratios
$m_\mathrm{d}=M_\mathrm{*}/M_\mathrm{T}$ are used: one with
$m_\mathrm{d}=0.02$ and minimum Toomre's $Q$ parameter $Q_\mathrm{min} \sim 1$
(hereafter referred to as the `low-$Q$' disc), the other with
$m_\mathrm{d}=0.01$ and $Q>1.3$ (hereafter referred to as the `high-$Q$'
  disc). As shown by previous works
    \citep[e.g.][]{vandervoort1970,romeo1992},  discs with finite thickness are more
    stable than razor-thin ones. Therefore, although for razor-thin discs, $Q\sim
    1$ leads to violent axisymmetric perturbations, in our simulations with 
    finitely thin discs such axisymmetric perturbations are weaker.  The radial profiles of $Q$ 
parameter for different discs and as a function of time are shown in
Fig.~\ref{fig:2} in Appendix~\ref{app:dssh}. The circular velocity and
velocity dispersion of the stars in the disc are worked out analytically based
on the density distribution of the system (for further detail, see \citet{springel2005}).

\subsection{Spherical Dark Matter Haloes}
\label{sec:hp}
Both Hernquist \citep{hernquist1990} and triaxial dark matter haloes derived
from it are used as our halo models. The Hernquist profile of a halo is
described by 
\begin{equation}
  \label{eq:da}
  \rho_{\rm HQ}(r)=\frac{M_{\rm dm}}{2\uppi}\frac{a}{r(r+a)^3}\,,
\end{equation}
where $a=30\,\mathrm{kpc}$ is a scale length factor that controls the
distribution of the mass. The potential of the Hernquist halo is given by
\begin{equation}
\label{eq:hq}
  \Phi_{\rm HQ}(r)=-\frac{GM_{\rm dm}}{r+a}\,.
\end{equation}
Here, the mass of the halo is
$M_\mathrm{dm}=M_\mathrm{T}(1-m_\mathrm{d})$. The virial radius of the halo,
$R_\mathrm{200}$, is set to be $160\,\mathrm{kpc}$.
\begin{table}
  \centering
  \caption{Halo models used in this work. $a$, $b$ and $c$ are major axis
  lengths of isodensity surface in the $x$, $y$ and $z$ direction. $p$ and $q$ are
the ratios of $b$ to $a$ and $c$ to $a$. The suffixes of the two
parameters, $0$ and $\infty$, indicate whether the parameter is for
the inner or the outer limit. In model  $T_{\rm O2}$,
the discs do not lie in the $x$--$y$ plane. The position of the disc plane is
defined with the first two Euler angles following the standard
  notation. In other words, it is decided  
by (1) setting coordinates $x'$--$y'$--$z'$ to initially coincide with $x$--$y$--$z$
coordinates, (2) rotating $x'$--$y'$--$z'$ coordinates along the $z'$ axis for
an angle $\alpha$, (3) rotating $x'$--$y'$--$z'$ coordinates along the $x'$ axis
for an angle $\beta$ and (4) putting the disc in the $x'$--$y'$ plane.}
  \begin{tabular}{rllllll}
    \hline
Model & $p_0$ & $q_0$& $p_\infty$ &
$q_\infty$& $\alpha$ & $\beta$ \\
    \hline
$S$ & 1 & 1 & 1 & 1 & 0 &0 \\
$T_1$ & 0.6 & 0.4 & 1 & 1& 0 &0 \\
$T_2$ & 0.95 & 0.85 & 0.6 & 0.5& 0 &0 \\
$T_3$ & 0.85 & 0.85 & 0.85 & 0.85& 0 &0 \\
$T_\mathrm{\rm O2}$ & 0.95 & 0.85 & 0.6 & 0.5& 0 & $\uppi/4$ \\
\hline
  \end{tabular}
\label{tab:1}
\end{table}

\subsection{Triaxial Dark Matter Haloes}

To derive a triaxial halo from a Hernquist halo, we followed
\citet{bowden2013} and added two low-order spherical harmonic terms to the
potential, i.e. 
  \begin{equation}
\label{eq:pot}
    \Phi(r,\theta,\phi)=\Phi_{\rm HQ}(r)+\Phi_\mathrm{T}(r,\theta,\phi)\,,
  \end{equation}
with the triaxial part $\Phi_\mathrm{T}(r,\theta,\phi)$ being
\begin{equation}
  \label{eq:pt}
\Phi_\mathrm{T}(r,\theta,\phi)=  4\uppi G \frac{\rho_1
      r_1^4 r}{(r+r_1)^3}Y_2^0(\theta,\phi)-4\uppi G \frac{\rho_2 r_2^4 r}{(r+r_2)^3}Y_2^2(\theta,\phi)\,,
\end{equation}
where $\Phi_{\rm HQ}$ is the potential of the Hernquist profile and
$Y_2^0(\theta,\phi)=\frac{3}{2}\cos^2\theta-\frac{1}{2}$,
$Y_2^2(\theta,\phi)=3\sin^2\theta\cos 2\phi$ are spherical harmonic functions.

Hence, 
\begin{multline}
\label{eq:den}
  \rho(r,\theta,\phi)=\rho_\mathrm{HQ}(r,\theta,\phi)-4\rho_1\frac{r_1^4(r_1^2+5r_1r+r^2)}{(r+r_1)^5r}Y_2^0(\theta,\phi)\\
+4\rho_2\frac{r_2^4(r_2^2+5r_2r+r^2)}{(r+r_2)^5r}Y_2^2(\theta,\phi).
\end{multline}

Here $\rho_1$, $r_1$, $\rho_2$ and $r_2$ are four free parameters that can be
used to control the shape of the halo in the inner and outer regions, i.e. the
ratio of major axis lengths of the isodensity surface in the $y$ and  $x$
direction at the inner and outer limit, $p_0=\lim_{r\rightarrow 0} b/a$ and
$p_\infty=\lim_{r\rightarrow \infty} b/a$ and similarly in the $z$ and $x$
direction, $q_0=\lim_{r\rightarrow 0} c/a$ and $q_\infty=\lim_{r\rightarrow
  \infty} c/a$. Once shape parameters $p_0$, $q_0$, $p_\infty$ and
  $q_\infty$ are given, one can calculate the corresponding $\rho_1$, $r_1$,
  $\rho_2$ and $r_2$ following \citet{bowden2013}. For most of the simulations the disc lies in the $x$--$y$ plane,
but we also ran several simulations in which the discs have a $45^\circ$ angle
to the $x$--$y$ plane. 

The names and parameters of all halo models are listed in Table~\ref{tab:1}.
The $S$ model is the original spherical Hernquist halo. $T_1$ model is
spherical outside and triaxial inside, while $T_2$ is the opposite. In $T_3$
model, the inner and outer limits of the major axis length ratios are
$p_0=q_0=p_\infty=q_\infty=0.85$. However the major axis length ratios $p$ and
$q$ are not constant throughout the halo. As shown in Fig.~\ref{fig:pp}, the
ratio $p$ is slightly lower than $0.85$ for $0<R<5R_\mathrm{S}$. This is due
to the fact that we are using an analytical model for the triaxial halo, which
only constrains the ratios of major axis lengths for the $r\rightarrow 0$ and
the $r\rightarrow\infty$ limits. The parameters for $T_2$ are chosen to be
comparable to the results of cosmological simulations described in
\citet{zemp2012}.

\section{Results}
\label{sec:results}
\subsection{Finite Resolution Effects}
\label{sec:finite-resol-effects}
We start by first investigating the finite resolution effects on the disc
properties by using an increasing number of star particles to represent the
disc. In total, we ran six simulations with the same $S$ halo, i.e. spherical
Hernquist halo without any triaxial terms, but with different disc mass ratio
$m_\mathrm{d}$ and different numbers of star particles in the disc. For two of
these simulations, we used an $m_\mathrm{d}=0.01$ disc with $10^6$ and $10^8$ star
particles. As mentioned before, in these disc models, Toomre's $Q$ parameter is
greater than $1.3$ throughout the disc. For the rest of the simulations,
$m_\mathrm{d}$ is set to $0.02$ and we use from $10^5$ to $10^8$ star
particles. For some regions in these discs, $Q \sim 1$, which means that the
swing amplification is strong.

\begin{figure}
  \centering
  \includegraphics[width=\linewidth]{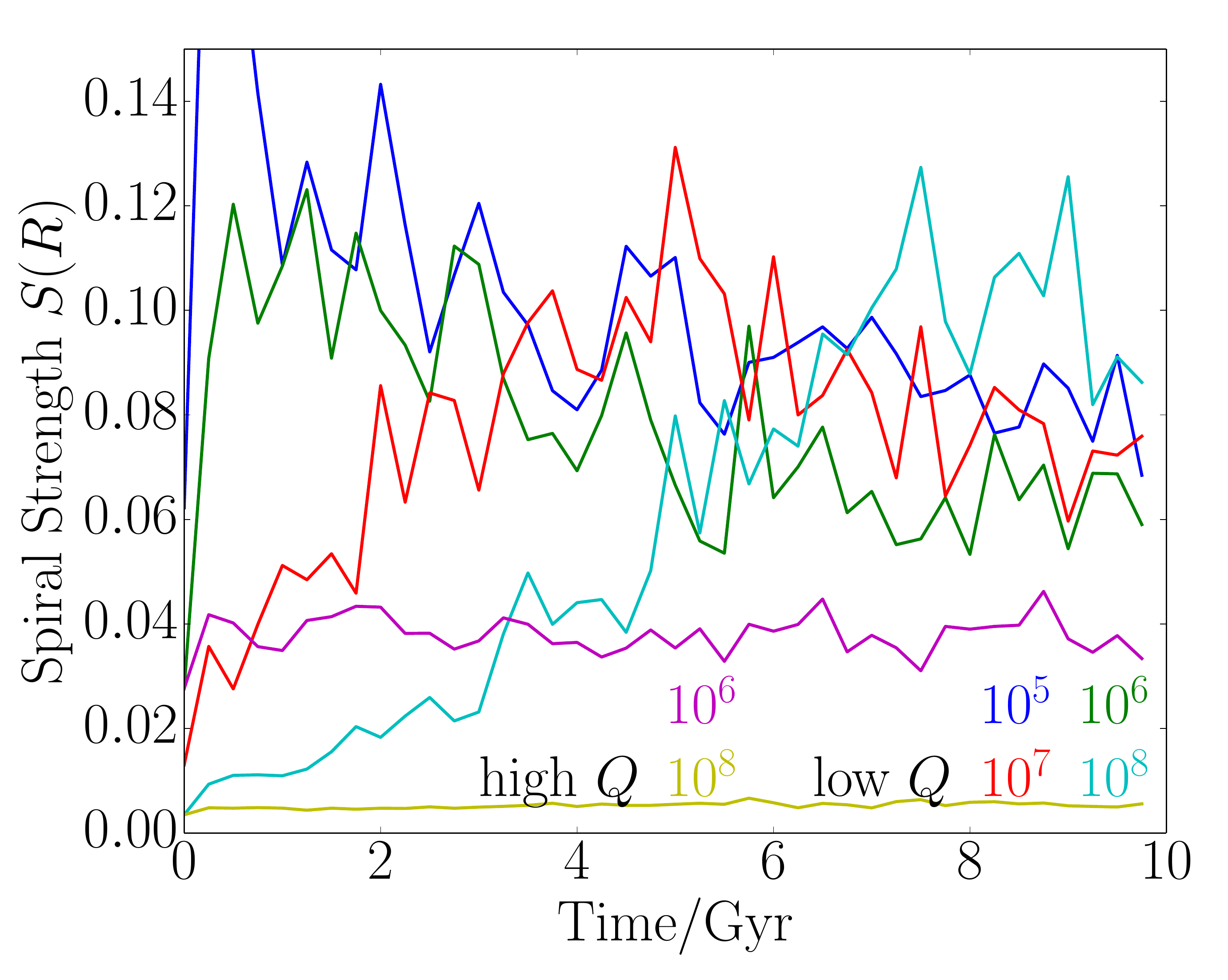}
  \caption{The growth of transient spiral structures with time for discs with
    $10^5$, $10^6$, $10^7$ and $10^8$ particles. Spiral strength $S(R)$ is
    defined in equations~\eqref{eq:1} and~\eqref{eq:fourier}, where all $m\le
    12$ modes are taken into account. Spiral structures grow in all four
    low-$Q$ discs, but not in high-$Q$ discs. For the high-$Q$ disc with
      $10^6$ particles, the strength is higher than that of $10^8$ particles due
    to higher Poisson noise, but does not grow over time. The scale time for growing
    spirals is longer for simulations with more star particles. For simulation
    with $10^5$--$10^7$ star particles, spiral strength decreases after
    reaching the peak value due to the spiral heating.}
\label{fig:svt}
\end{figure}

\begin{figure*}

      \includegraphics[width=.97\textwidth]{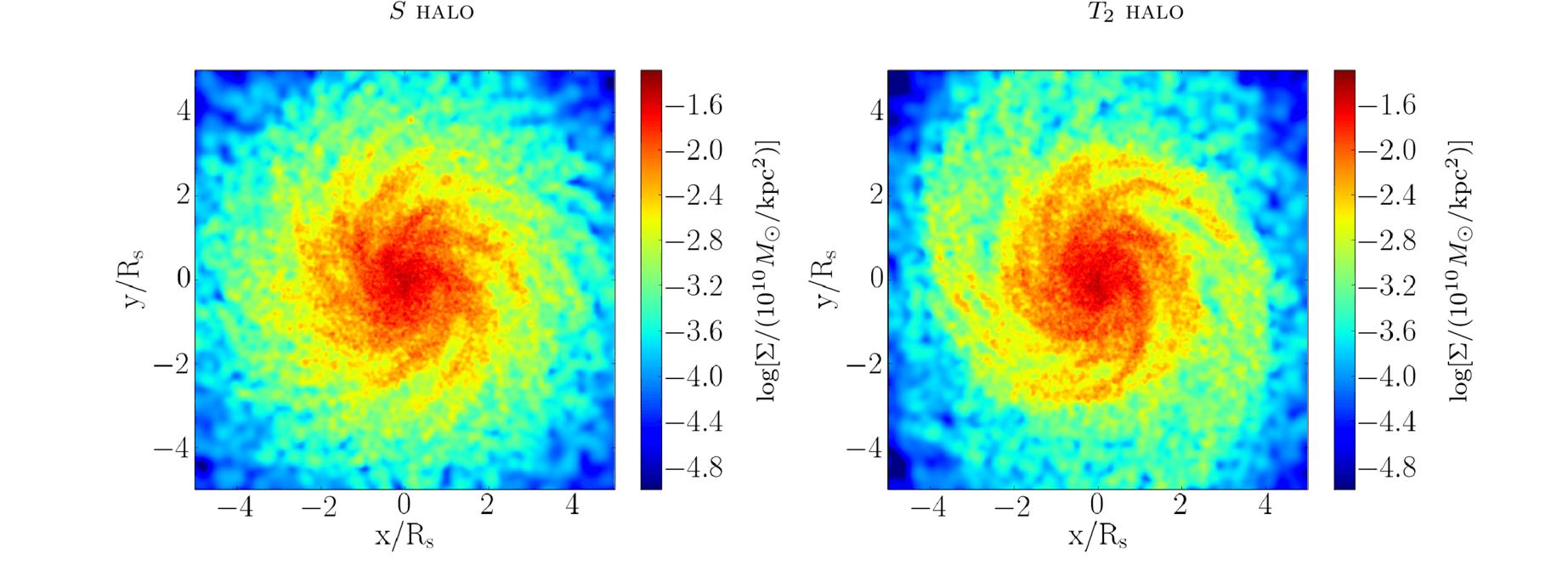}
    \caption{Surface density of a disc in the spherical and the $T_1$ triaxial
      halo at time $0.5\mathrm{Gyr}$. Values on the colour bar are
      $\mathrm{log}(\Sigma / 10^{10} \mathrm{M}_{\rm \odot}
      \mathrm{kpc}^{-2})$, where $\Sigma$ is the surface density. In both
      simulations the  disc has $10^5$ particles. The surface densities of the
      two discs are similar. This indicates that when the Poisson noise is
      significant, swing amplification of Poisson shot noise is the dominating factor over the
      influence of a triaxial halo.} 
\label{fig:53}
\end{figure*}

In simulations with low-$Q$ discs, transient spiral structures develop. As
shown in Fig.~\ref{fig:1} in Appendix~\ref{app:dssh}, these spiral
structures  are multi-armed, typically from 7-armed to 10-armed.  The strength
of these spiral structures at radius $R$ can be quantified with 
\begin{equation}
\label{eq:1}
  $S(R)$=\sqrt{\sum_{m=1}^{12} \left|\frac{\hat{\Sigma}_m(R)}{\hat{\Sigma}_0(R)}\right|^2}\,, 
\end{equation}
where $\hat{\Sigma}_m(R)$ is the Fourier transformation of the surface density
$\Sigma(R,\theta)$ of the disc at fixed radius $R$ along the azimuthal
coordinate $\theta$, i.e.
\begin{equation}
\label{eq:fourier}
  \hat{\Sigma}_m(R)=\frac{1}{2\uppi}\int_\mathrm{-\uppi}^{\uppi}\Sigma(R,\theta)\mathrm{e}^{-\mathrm{i}m\theta}\mathrm{d}\theta\,.
\end{equation}
Here we sum over $m=1$--$12$ to include the effect of all spiral
structures with $m\le 12$. Structures with higher $m$, which represent
smaller structures in the disc, are excluded because they are subject
to random noise in the disc. The evolution of  $S(R)$
over a time span of $10\,\mathrm{Gyr}$ at a fixed radius $R=2R_\mathrm{S}$ is
shown in Fig.~\ref{fig:svt}.

Strong spiral structures develop in all four simulations with low-$Q$
discs. The maximum spiral strength for these simulations is roughly the same
regardless of the number of star particles. However, simulations with a higher
number of particles take longer time to grow spiral structures, in
agreement with  the findings from \citet{Sellwood2012} and
\citet{Fujii2011}. This shows that self-gravitating discrete discs in our simulation can be well
  characterized by Balescu--Lenard equation, whereby the time-scale of
  structure growth is inversely
  proportional to the number of particles in the system \citep{fouvry2015d}. In particular, it takes more than $7\mathrm{Gyr}$ for the
simulation with $10^8$ star particles to fully develop spiral structures. Also
note that spiral strength in  simulations with $10^5$--$10^7$ particles
decreases after reaching the maximum, because of spiral heating. For
`dynamically hot' discs with $Q> 1.3$, the spiral strength is always
comparable to the initial value,
indicating that the swing amplification is weak, as expected.

To further study the resolution effects, we replace the spherical halo with
triaxial haloes described in Table~\ref{tab:1}. Fig.~\ref{fig:53} shows the
surface density of a $10^5$ particle disc in the $T_1$ halo at time
$t=0.5\,\mathrm{Gyr}$. $T_1$ halo is triaxial inside and spherical outside,
which means that the non-axisymmetric force introduced by this halo in the
central region of the disc is very high. However, the disc in the $T_1$ halo
still develops a multi-armed transient spiral structure, though the
distribution of the spiral arms is more asymmetric than the one in the spherical
halo. This indicates that when the Poisson noise in the disc is very high,
swing amplification of the disc dominates the whole process and external
torques cannot significantly influence the strength of the arms. To study
carefully the effects of triaxial haloes, swing amplification of Poisson noise
caused by finite resolution needs to be suppressed. For the rest of this
paper, we thus perform simulations with $10^8$ star particles, so that the
growth of transient spiral structures caused by swing amplification of the
Poisson noise in the initial conditions is sufficiently slow.

\subsection{Time-dependent Triaxial Haloes}
\label{sec:time-depend-triax}

\begin{figure*}
  \centering
    \includegraphics[width=\textwidth]{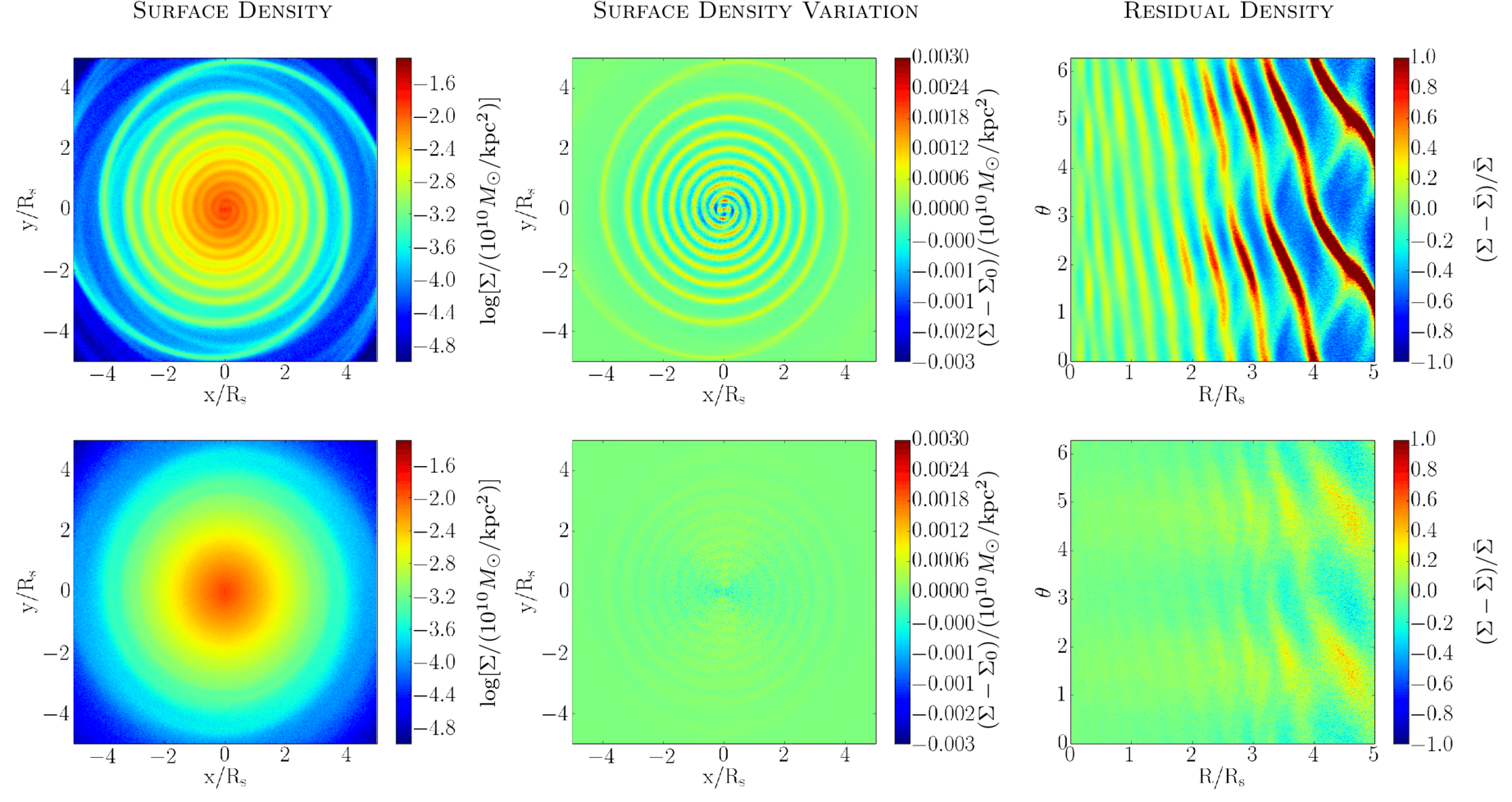}
  \caption{Surface density, i.e. $\mathrm{log}(\Sigma / 10^{10}
    \mathrm{M}_{\rm \odot} \mathrm{kpc}^{-2})$, its variation to the initial
    conditions and residual density of high-$Q$ discs in a simulation with
    $10^8$ star particles in a $T_2$ halo at $t=3\mathrm{Gyr}$. Top row:
    simulation with a static $T_2$ halo. Bottom row: halo is initially
    spherical and then gradually turns into a $T_2$ halo with a scale time of
    $\tau=1\mathrm{Gyr}$. In the simulation with a static $T_2$ halo, strong
    two-armed spiral structures form. The spiral structures keep winding up
    after their formation. By the time $t=3\mathrm{Gyr}$, spirals are tightly
    wound up. With triaxiality of the halo introduced gradually, very weak
    spiral structures form in the disc, indicating that the spiral structures
    found in the simulation with a static triaxial halo are caused by the
    sudden change of the halo shape.}
\label{fig:t21}
\end{figure*}

\begin{figure*}
  \centering
        \includegraphics[width=\textwidth]{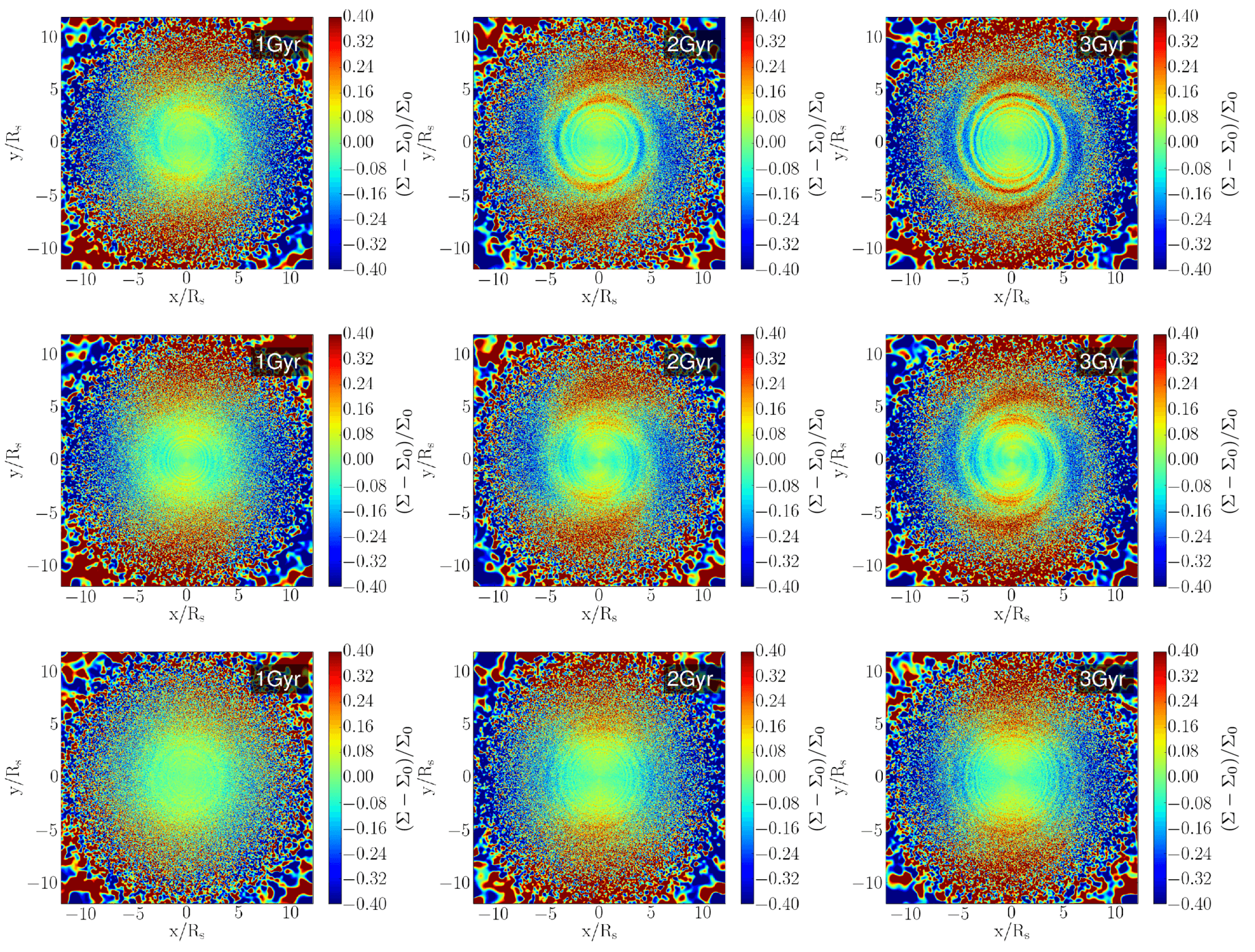}
  \caption{Discs in a $T_2$ halo with a gradually increasing triaxiality with
    time. The relative difference of surface density to the initial conditions
    is shown. Top: with self-gravity. Middle: without self-gravity. Bottom:
    with self-gravity and a smoother introduction of triaxiality in the 
  halo, i.e. $f_6(t)$ as defined in equation~\eqref{eq:gf}. With
  self-gravity in the disc turned off, it can be clearly seen that a
  two-armed spiral structure forms outside and grows inwards. However,
  for the simulation with self-gravity, this spiral pattern interferes with
  rings that form due to the change of halo profile. When the triaxiality of
  the halo is introduced in a smoother way, no prominent spiral structures can
  be found. This indicates that discs can survive in triaxial haloes without
  developing spiral structures if the shape of the halo changes smoothly
  enough.} 
  \label{fig:efoo}
\end{figure*}

The top row of Fig.~\ref{fig:t21} shows the surface density, density
variation with respect to the initial conditions and the residual density of
the high-$Q$ disc with $10^8$ star particles in the $T_2$ halo at
$3\mathrm{Gyr}$. Here residual density refers to the normalized density
difference to the average surface density over azimuthal coordinate,
\begin{equation}
\Sigma_\mathrm{res}(R,\phi)=\frac{\Sigma(R,\phi)-\bar{\Sigma}(R)}{\bar{\Sigma}(R)}\,,
\end{equation}
where $\bar{\Sigma}(R)$ is the average density at radius $R$,
i.e. $\bar{\Sigma}(R)=\frac{1}{2\uppi}\int_0^{2\uppi}\Sigma(R,\phi)\mathrm{d}\phi$. We use
a high-$Q$ disc to avoid the possible interference due to the swing
amplification of the disc, which we study later on in
Section~\ref{sec:swing-ampl-spir}. As shown in the top row of
Fig.~\ref{fig:t21} strong grand-design two-armed spiral structures
form. These spiral  structures are very sharp, tightly wound up and
global. Unlike the spiral structures formed due to swing amplification (see
Fig.~\ref{fig:1}), which only exist in the intermediate part of the disc
(i.e. for $1\lesssim R/R_\mathrm{S}\lesssim 3.5$), the spiral structures in
triaxial halo extend to the edge of the disc. This agrees with
\citet{dubinski2009} where an external torque due to the tumbling dark matter
halo drives the formation of  the spiral structures. However, it is unclear
whether these spiral structures are caused by the triaxial halo alone or by
the impulsive process of introducing triaxial haloes to the system with a disc
that is initially in equilibrium with a spherical halo.

To fully understand the formation mechanism of these grand-design spiral
structures, we run simulations in which the shape of the halo changes from
spherical to triaxial gradually, so that initially the disc is in the
equilibrium with the halo and no impulsive change to the system occurs. As
shown in equations~\eqref{eq:pot} and \eqref{eq:pt}, the analytic potential we
employed in our simulations consists of two parts: a spherical Hernquist
potential and a triaxial part. We hereby extend equation~\eqref{eq:pot} so
that the triaxial part of the halo can be turned on and off gradually with a
function $f(t)$, 
\begin{equation}
  \label{eq:hf}
   \Phi(r,\theta,\phi)=\Phi_{\rm HQ}(r,\theta,\phi)+f(t)\Phi_{\rm
     T}(r,\theta,\phi) \,.
\end{equation}
When $f(t)=0$, the dark matter halo is a purely spherical Hernquist halo,
while when $f(t)=1$, the dark matter halo is a fully developed triaxial halo. 

The bottom row of Fig.~\ref{fig:t21} shows the surface density and
the residual density of a simulation with a gradually introduced
$T_2$ halo. In this simulation the growth of the triaxial part of the
potential is set to 
\begin{equation}
\label{eq:of}
f(t)=1-\exp(-t/\tau_{\rm I})\,,
\end{equation}
where the scale time $\tau_\mathrm{I}=1\mathrm{Gyr}$. The time shown in
Fig.~\ref{fig:t21} is $3\mathrm{Gyr}$, when the triaxial transformation of
the halo is almost complete, with $f=0.95$. With the initially spherical halo
growing slowly to triaxial, as shown in the residual density map in the
rightmost panel, the spiral structure is very weak inside $3R_\mathrm{S}$,
while relatively weak two-armed structures develop in the outer region.

The fact that only weak spiral structures develop when the triaxiality of the
halo is introduced gradually indicates that triaxial halo alone does not
necessarily lead to spiral structures.  In fact, by comparing the orbital
period $\tau_\mathrm{O}$ of the stars  and the time-scale $\tau_\mathrm{I}$ of
introducing the halo, we find that the time-scales of the two processes
determine whether a strong spiral structure will develop. The orbital period
$\tau_\mathrm{O}$ is very small in the innermost regions of the disc. The
introduction of triaxiality has a much longer time-scale
$\tau_\mathrm{I}=1\mathrm{Gyr}$, which can be seen as an adiabatic
perturbation in the centre. $\tau_\mathrm{O}$ becomes longer at larger
radii. At $R=3R_\mathrm{S}$, $\tau_\mathrm{O}=0.315\mathrm{Gyr}$, which is
still shorter than the $\tau_\mathrm{I}$, but the introduction of the halo is
starting to make an effect. At about $R=5R_\mathrm{S}$, $\tau_\mathrm{O}\sim
0.5\mathrm{Gyr}$. Now introduction of the triaxial halo can no longer be
considered as adiabatic, and the change of halo shape starts to have a
significant effect on the disc.

The spiral structures formed in this simulation on larger scales are
shown in the first row of Fig.~\ref{fig:efoo}. Here the normalized density
difference to the initial density
\begin{equation}
  \overline{\Delta\Sigma}=\frac{\Sigma-\Sigma|_{t=0}}{\Sigma|_{t=0}}
\end{equation}
is shown so that the structures in outer regions where density is
significantly lower can be clearly seen. As shown in different columns
(which are for $t = 1, 2$ and $3\,\mathrm{Gyr}$), structures form at
larger radii and grow inwards. However, due to the changing of the halo
potential, ring structures also form in the disc\footnote{These are a
    numerical byproduct caused by the initial setup of stellar velocities
    which are not in perfect equilibrium with the changing halo potential, but
do not influence any of our results.} and to a certain extent
interfere with the spiral structures, making it  hard to distinguish genuine
spiral structures from rings. We therefore run another simulation with
self-gravity in the disc turned off, as shown in the middle row of
Fig.~\ref{fig:efoo}. Two-armed grand-design spiral structures still form in
this simulation, but rings do not interfere with spirals through self-gravity
anymore. Instead, they superimpose with each other so that the rings are now
seen as fine lines in each arm of the spiral structures. We therefore conclude
that rings, though forming easily as a response to the change of the halo
potential and modifying the shape of the spiral structures, are not essential
for the spiral forming process. 

\begin{figure}
  \centering
  \includegraphics[width=\linewidth]{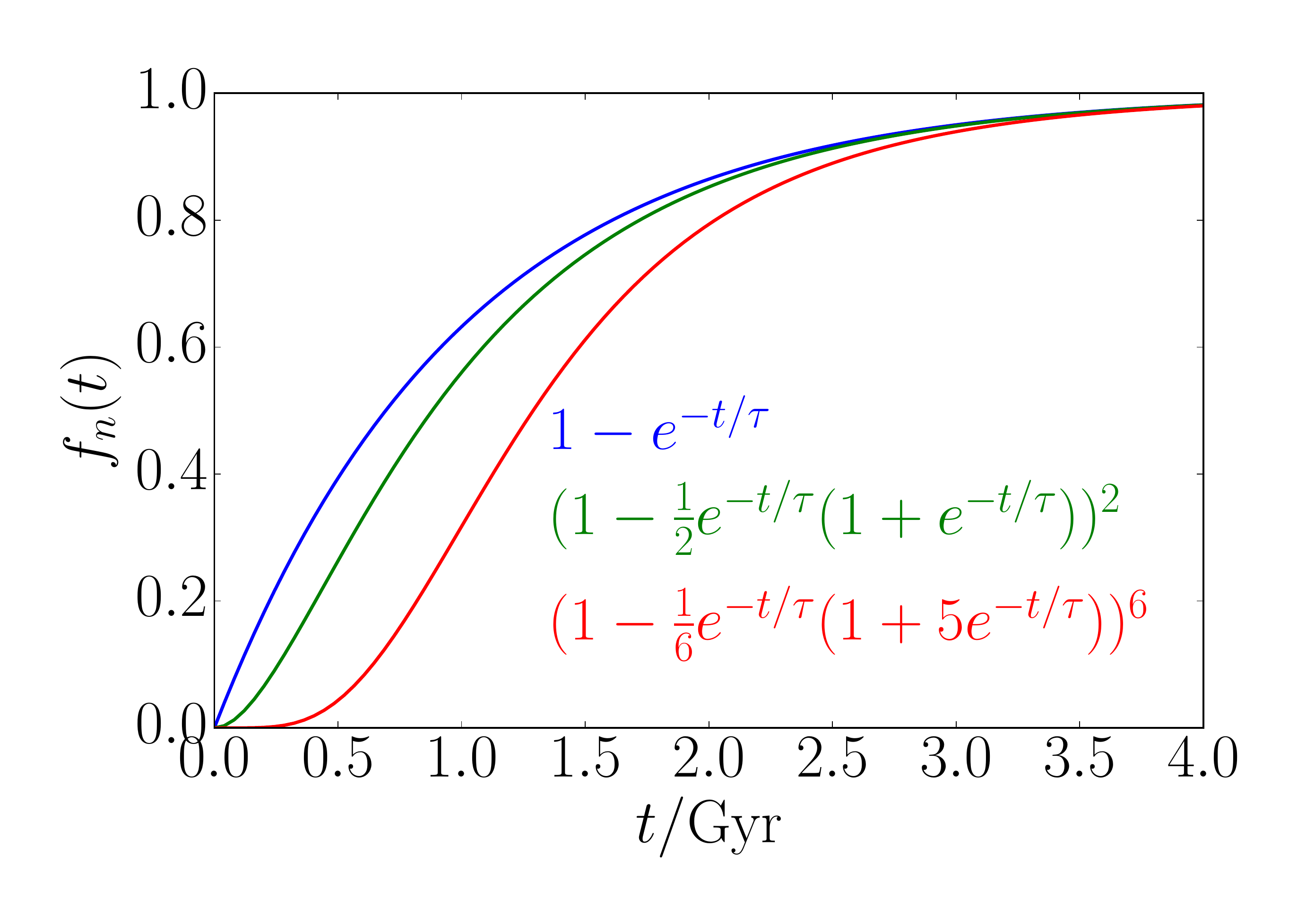}
  \caption{Time dependence of $f_n(t)$ for smoothing factor $n=1,2$ and
    $6$. $f_n(t)$ shows the same trend as $f_0(t)$ as $t\rightarrow
    \infty$. However higher $n$ indicates a smoother growth of $f_n$ at the
    beginning of the simulation.} 
  \label{fig:smooth}
\end{figure}

We explore this problem further with a smoother $f(t)$ function. We
generalized equation~\eqref{eq:of} to
\begin{equation}
  \label{eq:gf}
  f_n(t)=(1-\frac{1}{n} \mathrm{e}^{-t/\tau_\mathrm{I}}(1+(n-1)\mathrm{e}^{-t/\tau_\mathrm{I}}))^n\,,
\end{equation}
where $n$ is the smoothing factor and $f_1(t)$ falls to the original
form used in equation~\eqref{eq:of}. This generalization satisfies $f_n(0)=0$
and $f_n(t)=f_1(t)$ to the first order in $\mathrm{e}^{-t/\tau_\mathrm{I}}$ as
$t\rightarrow \infty$. As shown in Fig.~\ref{fig:smooth}, for a higher $n$,
the growth of $f(t)$ at the early stage of the simulation is smoother. In
practice we ran a simulation with $n=6$. Self-gravity is included in this
simulation as well. The result of this simulation is shown in the bottom row
of Fig.~\ref{fig:efoo}. No prominent spiral structures develop. This
demonstrates that the triaxial shape of a halo itself does not necessarily
lead to spiral structures. Rather, non-adiabatic change of the halo
shape, i.e. halo changing on a time-scale shorter than or comparable to
the orbital period of the stars, will cause grand-design spiral
structures to form.

\citet{Khoperskov2013} have reported that in their simulations similar
grand-design two-armed spiral structures form within a halo growing slowly
enough from a spherical one to a triaxial one with a time-scale more than four
times longer than the rotation periods of the stars at the outer part of the
disc. In contrast, we have shown in Fig.~\ref{fig:efoo} that with a
carefully chosen growth function $f(t)$ of the triaxiality of the halo, the
development of the spiral patterns is not necessary. We therefore conclude
that the spiral structures formed in their simulations are also due to the
fact that the introduction of the triaxiality is not adiabatic enough.

As also shown in the bottom row of Fig.~\ref{fig:efoo}, when the
  change of halo shape is adiabatic enough, the most prominent feature of the
  disc is its ellipticity. It has been shown that the ellipticity of the
  potential in the disc plane as well as the ellipticity of the disc itself can
  be constrained by the scatter of Tully--Fisher relation and by the photometry
  \citep{Franx1992,Debattista2008}. The potential
ellipticity is defined as $\epsilon=b_\Phi/a_\Phi$ where $a_\Phi$ and
  $b_\Phi$ are the length of major and minor axis of the surface of a constant
  potential. In our $T_2$ model, it varies from $\epsilon=0.055$ to $0.075$ in the disc plane from the inside to
the outside. The ellipticity of the disc, defined similarly by
  contours of disc surface density, is $\epsilon_\mathrm{D}=0.05$. Both of the
ellipticities
are consistent with $\epsilon<0.1$ and $\epsilon_D=0.06$ as suggested by \citet{Franx1992}.

\begin{figure*}
  \centering
  \includegraphics[width=\linewidth]{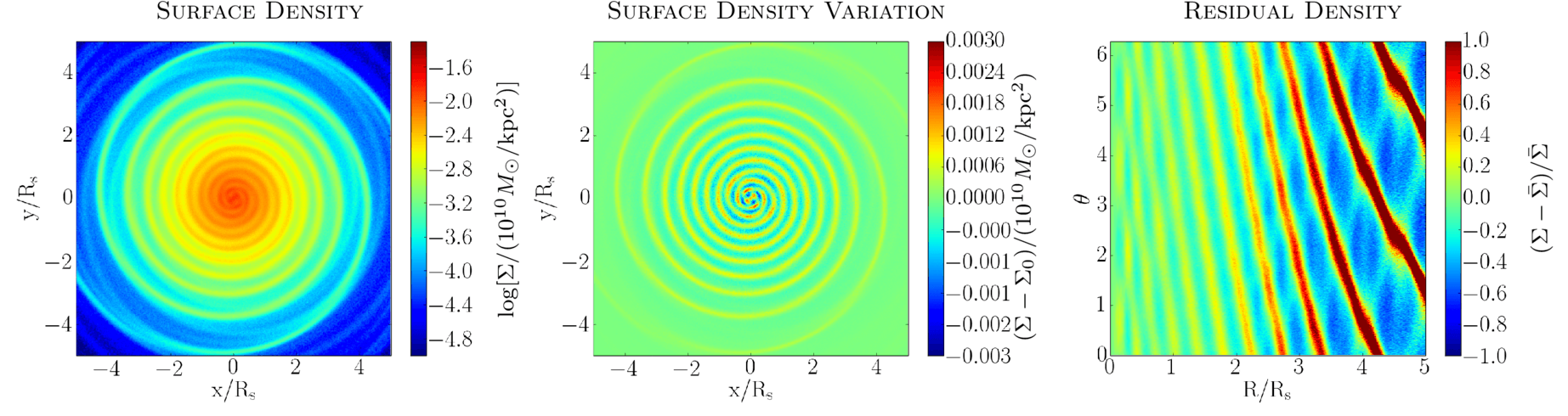}
  \caption{Simulated discs at $t=3\,\mathrm{Gyr}$ in a halo that has a
    time-dependent triaxiality. Halo starts as the $T_2$ halo, but is quickly
    turned into a spherical halo with a scale time of
    $\tau=0.2\mathrm{Gyr}$. The grand-design spiral structures still exist for
    a long time, indicating that once formed, grand-design spiral patterns can
    survive in a spherical halo.}
  \label{fig:efo}
\end{figure*}

To understand if triaxial haloes are needed for these grand-design
spiral structures to survive for longer periods of time, we also run a
simulation with the halo shape turning from triaxial to spherical abruptly,
with
\begin{equation} 
  f(t)=\exp(-(t/\tau_{\rm D})^2)\,,
\end{equation}
with a time-scale $\tau_{\rm D}=0.2\mathrm{Gyr}$. In this simulation strong
spiral structures form at the beginning of the simulation almost instantly, as
expected. We run simulation further in time and find that the spiral
structures can persist for much longer time. The surface density, its
variation to the initial conditions and the residual density of the disc at
$t=3\mathrm{Gyr}$ are shown in Fig.~\ref{fig:efo}.  The halo is
extremely close to spherical as early as $t=0.5\,\mathrm{Gyr}$, with $f\sim 10^{-3}$. Spiral structures
survive in this spherical halo for at least $2\mathrm{Gyr}$ and remain
strong. Therefore, triaxial haloes, proven above not necessary for forming
spiral structures, are not needed for persisting spiral structures either.

It is interesting to note that  \citet{toomre1981} studied properties of a
disc galaxy with an external torque turned on and off rapidly. Though both in
Toomre's and in our simulations, two-armed spiral structures form, they are
different in at least two ways. First, in \citet{toomre1981}, the Toomre's $Q$
parameter and the wavelength ratio $X$ are both in the range that favours the
swing amplification of $m=2$ modes, while in our simulation, for $m=2$ the
wavelength ratio $X$ is too high for swing amplification to  become significant. Actually in
our simulations the modes for the $X$ to fall into the range $1<X<3$ that
favours the swing amplification require $7\lesssim m\lesssim 10$, which are
the typical number of arms for transient spiral structures seen in
Fig.~\ref{fig:1}. Therefore, though the swing amplification plays an
important role in the development of the grand-design spiral structures in
\citet{toomre1981}, it plays a minor role in our simulations. In fact, as
shown later in Section~\ref{sec:swing-ampl-spir}, the swing amplification may
destroy the  grand-design spiral structures at later times. Secondly,
\citet{toomre1981} found that the grand-design spiral structures in their
simulation decay after several rotational periods, while in our simulations
the grand-design spiral structures can survive for a longer time. While disc
properties and simulation methodology in our work are quite different, this
indicates that the formation mechanism of the spiral structures may be
different in \citet{toomre1981} and our simulations.  In fact, the spiral
structures in our simulations can be well explained by the kinematic density
wave theory, as shown in Section \ref{sec:swing-ampl-spir}.

\subsection{Time-dependent Triaxial Haloes from Cosmological Simulations}
\label{sec:triax-halo-cosm}

\begin{figure*}
  \centering
\begin{minipage}[t]{.45\linewidth}
  \centering \sc Triaxiality Profile of the First Model
\end{minipage}\begin{minipage}[t]{.45\linewidth}
  \centering \sc Triaxiality Profile of the Second Model
\end{minipage}
  \includegraphics[width=.45\linewidth]{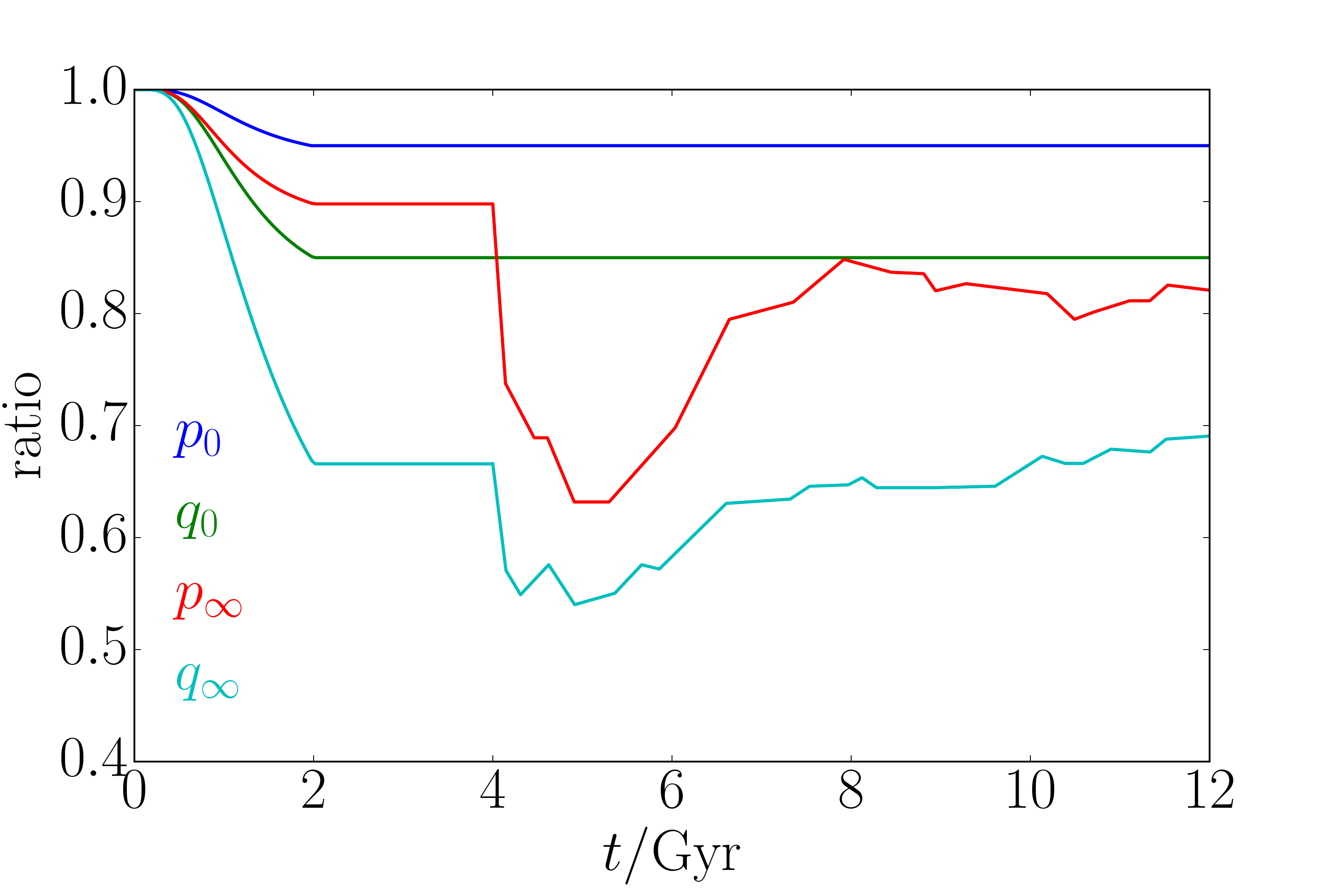}\includegraphics[width=.45\linewidth]{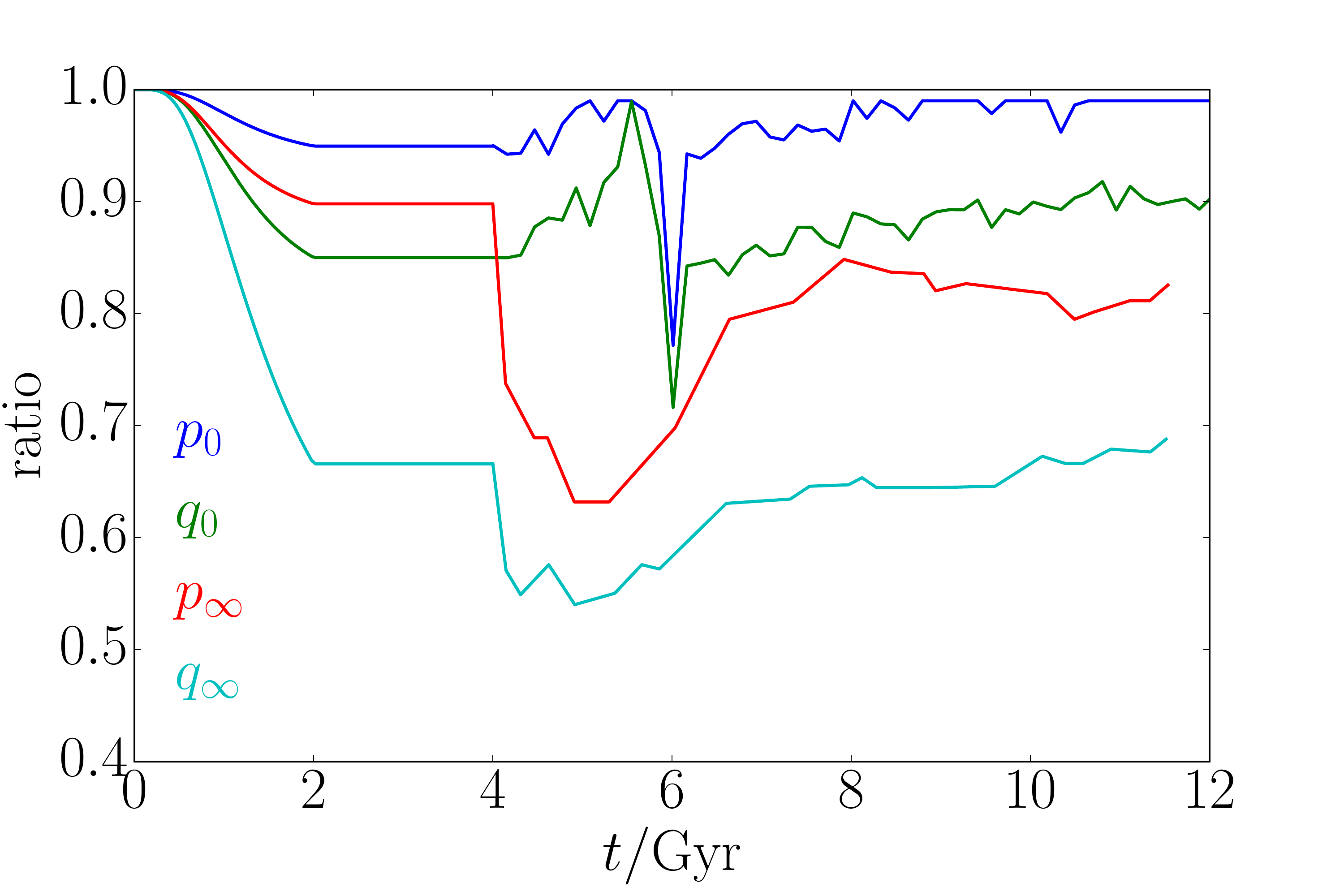}
  \caption{Time dependence of the major axial ratios of the halo
      isodensity 
      surface. For both models, from 0 to $2\,\mathrm{Gyr}$, the halo grows from
    spherical to triaxial smoothly. At $2\,\mathrm{Gyr}$, the triaxiality of the
    outer halo is the same as that of the Aq-B-4 halo at $5.4\,\mathrm{Gyr}$
    of the original simulation, as shown
    in fig. 5 of 
    \citet{Vera-ciro2011}. The halo is kept static until
    $4\,\mathrm{Gyr}$, allowing the disc to fully relax. From
    $4\,\mathrm{Gyr}$, the outer side of
    the halo evolves similarly to the Aq-B-4 halo starting from
    $5.4\,\mathrm{Gyr}$ in 
    the Aquarius simulation, while the limits of the major axial ratios as
    $r\rightarrow 0$ remain constant for the first model (left-hand panel), but evolve
      following a recalibration based on the Aq-B-4 simulation for the second model (right-hand panel).} 
  \label{fig:realhalo}
\end{figure*}

As shown in the previous section, spiral structures only form when the
axisymmetric change of halo potential is rapid enough. It is natural to ask
whether this process is possible in a realistic halo. In fact,
\citet{Vera-ciro2011} found that haloes in self-consistent cosmological
simulations like 
the Aquarius runs \citep{Springel2008} can have very rapid change in
triaxiality over time even when not experiencing very violent major mergers.
 To directly explore the impact of such realistic triaxiality
changes on stellar discs, we set up two isolated halo
models. For both models, the axial ratios $p=b/a$ and $q=c/a$ at the outer side of the halo
follow Aquarius halo Aq-B-4 [see fig. 5 in \citet{Vera-ciro2011} at $\sim
5.4\,\mathrm{Gyr}$ of their simulated time]. The
  first model we consider has constant axial ratios with time in the central region. The limit of triaxiality as
$r\rightarrow 0$ is set to be $p_0=b/a|_{r\rightarrow 0} = 0.95$ and
$q_0=c/a|_{r\rightarrow 0} = 0.85$ similar to the $T_2$ model 
\citep[following][]{zemp2012}, to account for the likely impact of baryonic matter. 

This
  model is rather conservative as it does not account for any 
  time variation of triaxiality in the innermost halo regions. Thus, for the second
  model, we directly calculate the time evolution of the inner triaxiality of
  the Aq-B-4 halo. Due to the possible influence of a baryonic component, which is not captured by the
Aquarius simulation (as it contains dark matter only), we expect that the
average inner triaxiality will be smaller, but that the time fluctuations of
triaxiality triggered by mergers, for example, will be comparable as predicted by
dark matter-only cosmological simulations. We hence impose inner halo
triaxiality fluctuations as directly evaluated from the Aq-B-4 halo on top of more
realistic $p_0=b/a|_{r\rightarrow 0} = 0.95$ and $q_0=c/a|_{r\rightarrow 0} =
0.85$ values, rather than taking the absolute Aq-B-4 inner triaxiality average
values, which are in the range of $\sim 0.6$. With this ``recalibration'' our second model
more closely reflects the possible inner triaxiality time evolution as would
be obtained by very high resolution Aq-B-4 simulation which would include
baryons. Thus the impact of halo triaxiality on the stellar disc should be
more realistic than in our first model.

The time dependence of halo triaxiality of our models is shown in Fig.~\ref{fig:realhalo}. We start from a spherical halo and
change the triaxiality of the halo adiabatically to the configuration
resembling the Aq-B-4 halo at $5.4\,\mathrm{Gyr}$ of the original simulation, and
let the disc relax for $2\,\mathrm{Gyr}$ before 
the outer halo triaxiality starts to evolve similarly to the Aq-B-4 halo onwards in
time. In the left-hand panel of Fig.~\ref{fig:realhalo} we show our first
model where the inner halo triaxiality is kept constant, while in the
right-hand panel, corresponding to our second model, realistic cosmological
fluctuations of inner halo triaxiality are imposed (as measured from the
Aq-B-4 simulation directly) on top of the same constant values as in the first model.

\begin{figure*}
  \centering
      \includegraphics[width=\linewidth]{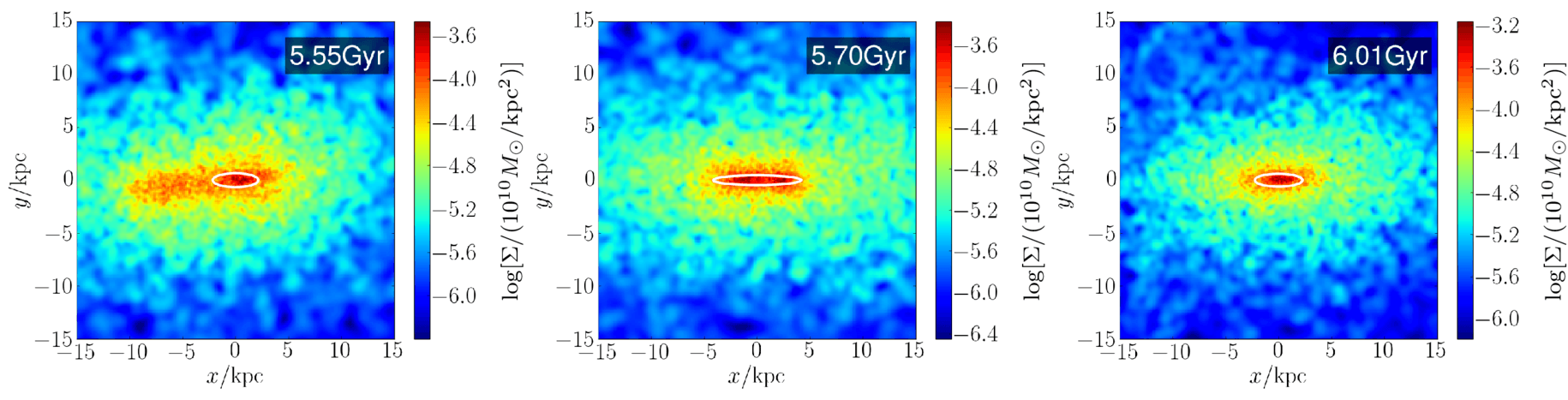}
  \caption{Time sequence of the central Aq-B-4 halo from $5.55$
  to $6.01\,\mathrm{Gyr}$. The panels show the projected density of the dark
    matter halo.
  White ellipse represents the fit to the halo shape at
  $\sqrt[3]{abc}=1\,\mathrm{kpc}$, where $a, b$ and $c$ are the lengths of the three
  axes. A satellite merges with the  Aq-B-4 halo during
  this period, which leads to a temporary but abrupt halo shape change. } 
  \label{fig:aq-b-halo}
\end{figure*}

\begin{figure*}
  \centering
  \includegraphics[width=\linewidth]{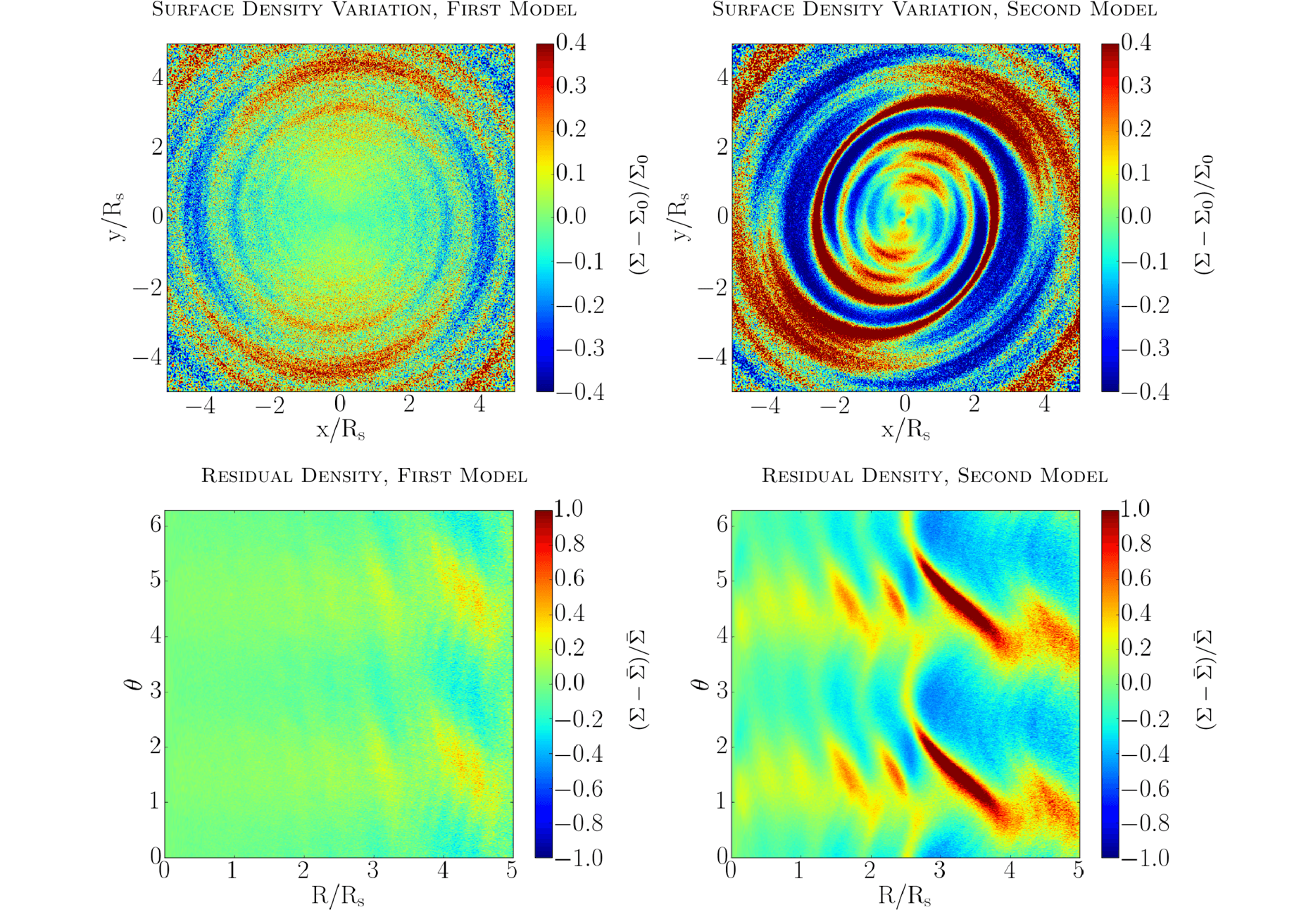}
  \caption{Surface density difference to the initial conditions (top) and
      residual densities (bottom) at
    $t=6\,\mathrm{Gyr}$ of discs evolved in  simulations shown in
      Fig.~\ref{fig:realhalo}. The halo starts to behave like Aq-B-4 halo
      after $t=4\,\mathrm{Gyr}$ in this simulation. At $t=6\,\mathrm{Gyr}$,
      clear spiral structures already develop for both simulations,
      while for the second model with varying inner triaxiality, the spiral
      strength is  larger, and spirals form further in.}
  \label{fig:realhalo-surf}
\end{figure*}

\begin{figure*}
  \centering
\begin{minipage}[t]{.55\linewidth}
  \centering \sc Ratio of Major Axes
\end{minipage}\begin{minipage}[t]{.5\linewidth}
  \centering \sc Torque Strength
\end{minipage}
  \includegraphics[width=.5\linewidth]{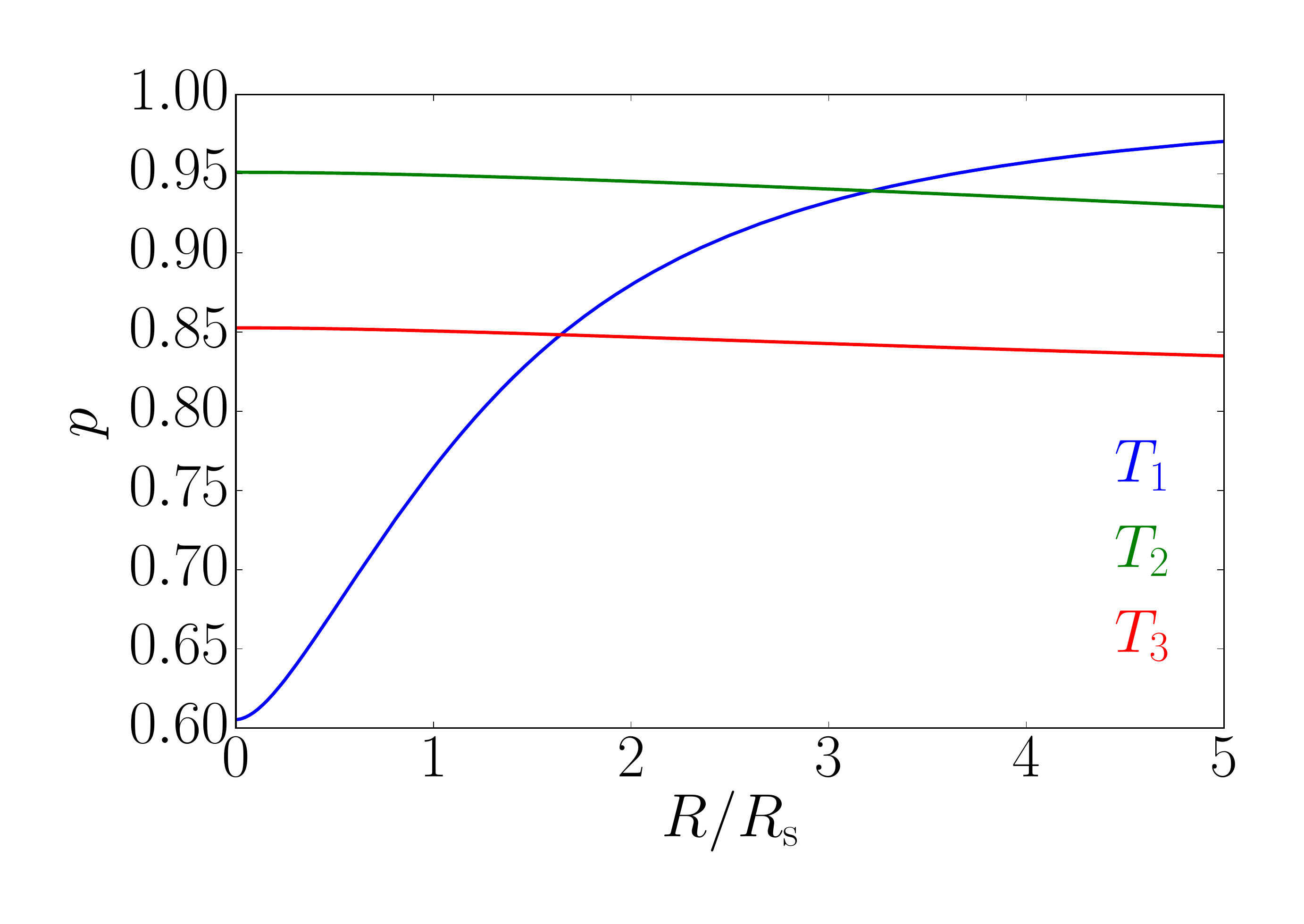}\includegraphics[width=.5\linewidth]{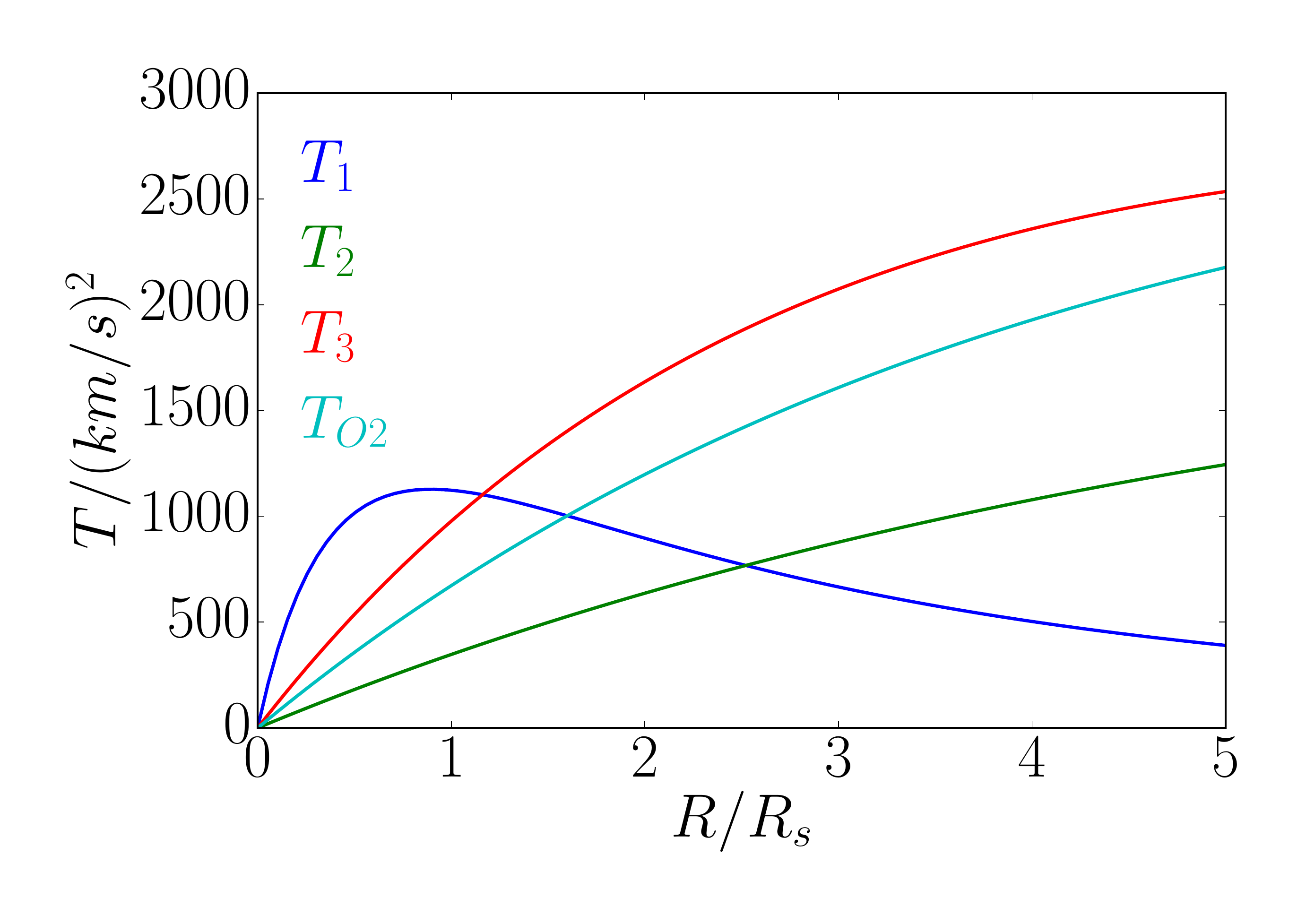}
  \caption{Shape and torque strength of the dark matter halo models as a
    function of $R/R_\mathrm{S}$. Left: ratio of major axes, $p=b/a$, as a
    function of $R/R_\mathrm{S}$ for halo models $T_1$, $T_2$ and $T_3$ listed
    in Table~\ref{tab:1}. Right:  absolute value of the torque perpendicular
    to the disc plane that a star with azimuthal coordinate $\theta=\uppi/4$ and
    radius $R$ feels for simulations with halo models $T_1$, $T_2$, $T_3$ and
    $T_\mathrm{O2}$. The simulation with halo model $T_\mathrm{O2}$ is
    discussed later in Appendix~\ref{sec:discs-misal-triax}.}
  \label{fig:pp}
\end{figure*}

A prominent feature in the right-hand panel of
  Fig.~\ref{fig:realhalo} is a sharp dip at around $6\,\mathrm{Gyr}$. A
  minor merger is found to be the cause for such a sharp change, as illustrated 
  in Fig.~\ref{fig:aq-b-halo}. We trace this merger event
  backwards in time, and find that the satellite that causes this merger event has
  initially $\sim 23\%$ of the mass of the Aq-B-4 halo. As it inspirals towards the centre of the Aq-B-4
  halo, which lasts for about $3\,\mathrm{Gyr}$, it loses most of its mass gradually due to dynamical friction and tidal
  stripping. None the less, the core of the satellite
  disrupts the centre of the Aq-B-4 halo significantly as it finally merges with it.

For both models, as intended the disc shows no sign of spiral structures in the growing
  and relaxing phase over the first $4\,\mathrm{Gyr}$. However, when our halo
  triaxiality starts to evolve like that of the Aq-B-4 halo,
clear spiral structures form in about $2\, \mathrm{Gyr}$, as shown
in Fig.~\ref{fig:realhalo-surf}. The spiral structures persist and sharpen
over time for at least another $3$--$4\, \mathrm{Gyr}$. The morphology of the spiral structures is
very similar to those shown in Section~\ref{sec:time-depend-triax}.
Generally, the spiral structures of the second model with changing inner
  triaxiality are much stronger than that of the first model. Also, for the second model the spiral strength is
high all through the disc\footnote{The evolution of the
spiral strength of the second model will be discussed later in Section~\ref{sec:formation-mechanism}.}, while for the first model, the spiral
structures are stronger in the outer part of the disc, which is expected since
the inner limit of triaxiality is fixed. None the less, the spiral
  strength of the first model is still considerable as the relative surface
  density fluctuations reach $40\%$. We therefore conclude
that in a realistic, cosmologically motivated case, changes in the halo
triaxiality are abrupt enough to cause spiral structures, even when we assume
a conservative evolution for the innermost halo. In reality, due to minor mergers, the inner shape
of the halo may change in a way similar to our second model, therefore
leading to strong spiral structures similar to the right-hand panels of Fig.~\ref{fig:realhalo-surf}. More detailed investigation of this issue, which depends on
the complex interplay of baryons and inner halo triaxiality fluctuations with
time is beyond the scope of this work and is left for a future study.

It is also possible that in reality the disc is misaligned with respect
  to  the halo's major
plane. We explore this in Appendix~\ref{sec:discs-misal-triax} and find that
the misaligned disc  no longer
stays in a plane as it evolves. Integral shaped warps form in the
disc. Two-armed spiral structures still form and are similar to those formed
in discs that lie in the $x$--$y$ plane of the triaxial haloes. However, the outer
parts of the spiral structure are distorted, due to the fact that they are no
longer in the disc plane.

\subsection{Dependence of Spiral Strength on the Halo Shape}
\label{sec:depend-spir-strength}

\begin{figure*}
\centering
\includegraphics[width=\linewidth]{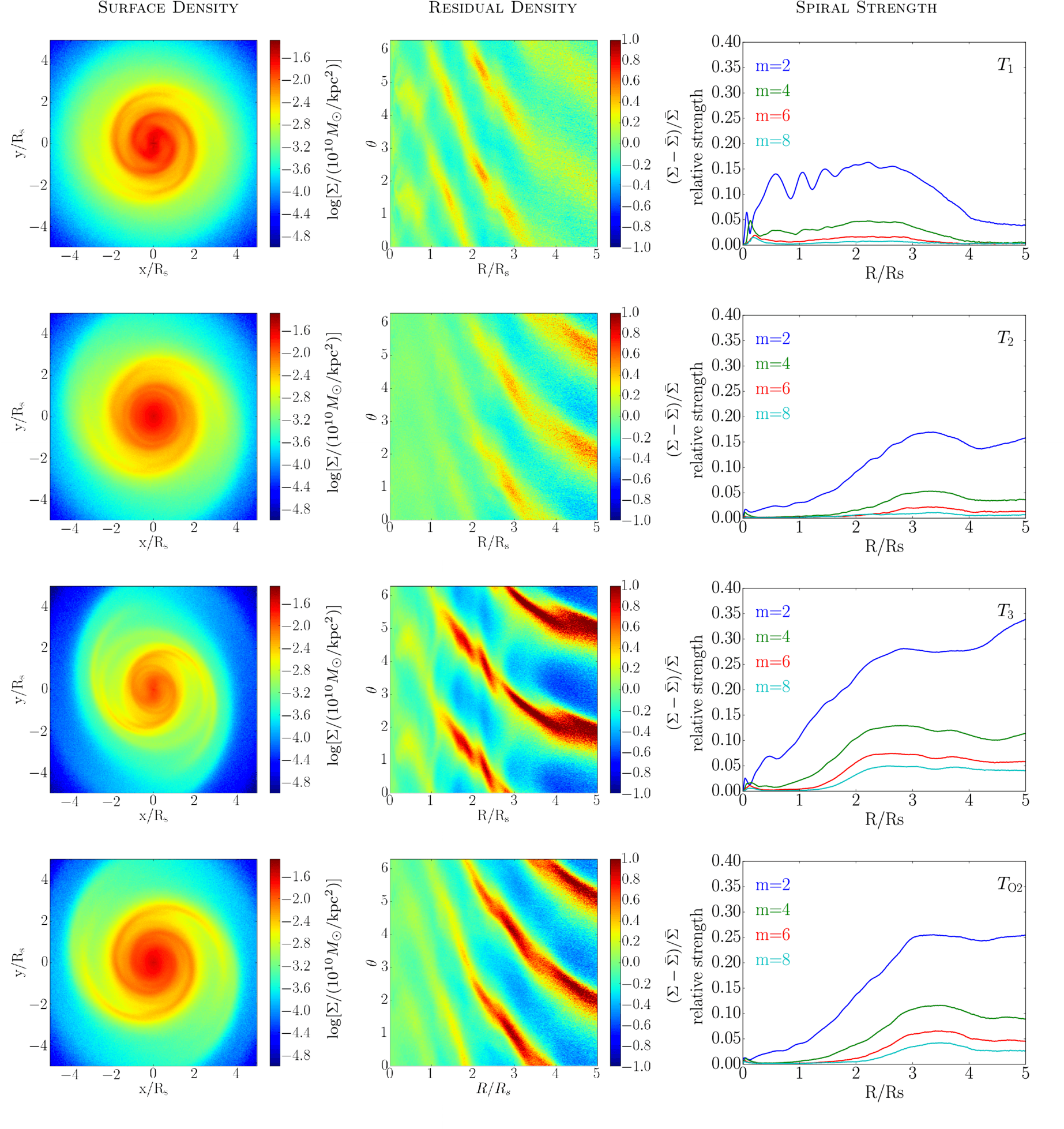}
  \caption{Surface density, residual density and spiral structure strength of
    discs in simulations with $T_1$, $T_2$, $T_3$ and $T_\mathrm{O2}$ haloes
    (from top to bottom). Left column: surface density in Cartesian
    coordinates at time $t=0.5\,\mathrm{Gyr}$. Central column: residual
    density in polar coordinates at time $t=0.5\,\mathrm{Gyr}$. Right column:
    spiral strength for different modes averaged over a time interval from
    $0.35$ to $0.85\,\mathrm{Gyr}$. This time interval in chosen
    so that the spiral structures are fully developed in the disc. Note that
    spiral strengths with the odd $m$ modes are not shown because they are
    much weaker. For the halo that is more triaxial inside ($T_1$), the spiral
    structure is stronger inside, while for the halo that is more triaxial
    outside ($T_2$), the spiral structure is also stronger outside. For $T_3$
    halo which has $b/a \sim c/a \sim 0.85$, the spiral structure strength is
    generally  higher than the other two simulations, expect for the central
    region where the spiral structure strength is lower than that of the
    $T_1$ halo. For the $T_\mathrm{O2}$ halo, the disc plane is  determined by
    the direction of the total angular momentum, as discussed later in
    Appendix~\ref{sec:discs-misal-triax}.}
  \label{fig:3}
\end{figure*}

\begin{figure*}
        \includegraphics[width=\textwidth]{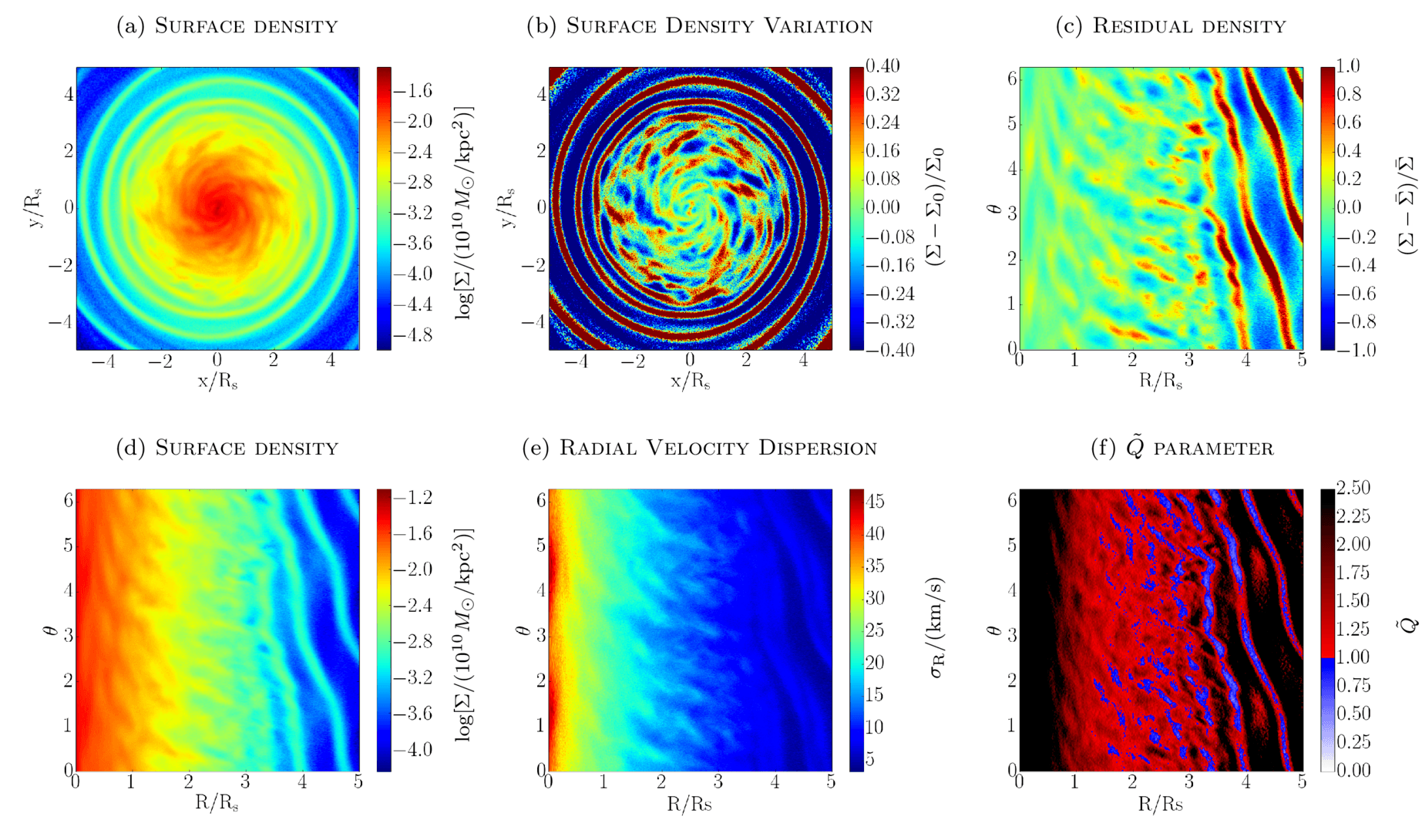}

    \caption{Maps of various quantities of a low-$Q$ stellar disc in the
      $T_2$ halo 
      at time $3\,\mathrm{Gyr}$. Panel (a) shows the surface density in Cartesian
      coordinates. To show the structures throughout the disc more clearly, we
      subtracted the initial surface density $\Sigma_0$ from the surface
      density $\Sigma$ at $3\mathrm{Gyr}$ and then divided it by
      $\Sigma_0$. The result is shown in panel (b). Panel (c) shows the residual density
      map in polar coordinates. In panel (d), (e) and (f), surface density,
      dispersion of the velocity in the radial direction,  $\sigma_\mathrm{R}$,
      and $\tilde{Q}$ parameter are shown in polar coordinates. Here
      $\tilde{Q}(R,\theta)=\frac{\sigma_R\kappa}{3.36 \,G \Sigma}$. Toomre's $Q$
      parameter is the angular average of $\tilde{Q}$. $\tilde{Q}$ quantifies the local
      stability of the disc.
      Two different types of
      spiral structures co-exist in the disc. Outside $3.5R_\mathrm{S}$ is the
      grand-design two-armed spiral structure caused by the triaxial halo,
      while inside $3.5R_\mathrm{S}$ there are more arms and they are all
      transient.}
    \label{fig:lt}
  \end{figure*}

We performed three additional simulations to further study the influence of
different halo  shapes on the spiral structure strength. In these simulations,
we employ low-$Q$ discs and various static triaxial dark matter
haloes. The disc is originally in equilibrium with a spherical halo. An abrupt
introduction of the triaxiality in the halo is the cause of spiral
structures, as shown in Section~\ref{sec:time-depend-triax}.  

Three halo models, $T_1$, $T_2$ and $T_3$, introduced in Table~\ref{tab:1},
are used in these simulations. The ratio of major axes, $p = b / a$, of these
haloes at different radii is shown in the left-hand panel of Fig.~\ref{fig:pp}.
$T_1$ model is more triaxial inside, $T_2$ model is more triaxial outside, and
the ratio of the major axis of the halo in the $T_3$ model is equal at the
innermost and the outermost limits, and slightly lower in between.  In all
three simulations, spiral structures develop instantly in response to the
abrupt change of the halo shape from the spherical one. Surface density
$\Sigma$, residual density $\Sigma_\mathrm{res}$ and the relative spiral
strength of different $m$ modes, $S_m$, of the discs in these simulations are
shown in Fig.~\ref{fig:3}. Here the relative spiral strength $S_m$ is
defined as
\begin{equation}
  \label{eq:spiralstrength}
  S_m(R)=\left | \hat{\Sigma}_m(R)/\hat{\Sigma}_0(R)\right |\,,
\end{equation}
where $\hat{\Sigma}_m(R)$ is the Fourier transformation of the surface density
of the disc at radius $R$ along the azimuthal coordinate, as defined in
equation~\eqref{eq:fourier}. In practice, the spiral strength fluctuates over
time at a given radius. To show the mean trend of the strength with respect to
the radius, we therefore plot the averaged spiral strength over a time
interval from $0.35$ to $0.85\,\mathrm{Gyr}$.

As shown in the top three rows of Fig.~\ref{fig:3}, two-armed spiral
structures form in all three simulations. The spiral strength for $m = 2, 4,
6$ and $8$ has a similar radial profile within each simulation, indicating
that the spiral strength measured for higher $m$ modes is largely due to the
two-armed spiral structures. However, for different simulations, the relative
spiral strength, $S_m$, varies differently with radius $R$. Generally, spiral
strength depends on the halo triaxiality.  This can be understood by comparing
the torque generated by the triaxial  halo, as shown in the right-hand panel of
Fig.~\ref{fig:pp}. The torque and the strength of the spiral  structures
show almost identical trends as a function of radius for all  three
simulations, i.e. by comparing different simulations, one can find that the
strength of the spiral structures at a given radius is higher for the
simulation where the torque at that radius is stronger, and vice versa.

\subsection{Swing Amplification of Spiral Fragments}
\label{sec:swing-ampl-spir}
As shown in Fig.~\ref{fig:svt}, for a low-$Q$ disc with $10^8$ star
particles located in a spherical halo, transient spiral structures develop in
several $\mathrm{Gyr}$ due to swing amplification. To explore if this process
will also take place in simulations with a triaxial halo, we run the
simulation with a $T_2$ halo for a longer time.

As shown previously in Fig.~\ref{fig:3}, at first only two-armed spiral
structures develop in the disc. However after some time transient multi-armed
spiral structures gradually form in the central region of the disc. Various
properties of the disc in this simulation at $t=3\,\mathrm{Gyr}$ are shown in
Fig.~\ref{fig:lt}. It can be seen from the surface density of the disc that
two kinds of spiral structures co-exist in the disc, with transient spiral arm
structures dominating the inner region of the disc, and two-armed grand-design
spiral structures taking up the outer region.

Generally, the swing amplification is
strong when the $Q$ parameter is close to unity. Fig.~\ref{fig:lt} shows that regions with low
$\tilde{Q}(R,\theta)=\frac{\sigma_R\kappa}{3.36 \,G \Sigma}$ value (note that Toomre's $Q$
      parameter is the angular average of $\tilde{Q}$) lie perfectly within
the spiral structures. This indicates that the spiral 
arms, with their higher density, can attract more stars into the spiral
arms, hence magnifying the strength of the spiral structures. Also, in the
part of the disc with $R_\mathrm{S}<R<3.5R_\mathrm{S}$, the $\tilde{Q}$ parameter is
generally slightly higher than $1$ outside of the spiral structures.  This
explains why transient spiral structures form in this region due to swing
amplification, while for $R > 3.5R_\mathrm{S}$ azimuthally averaged $\tilde{Q}$ value
is significantly higher than $1$.
\begin{figure*}
  \centering
  \includegraphics[width=\linewidth]{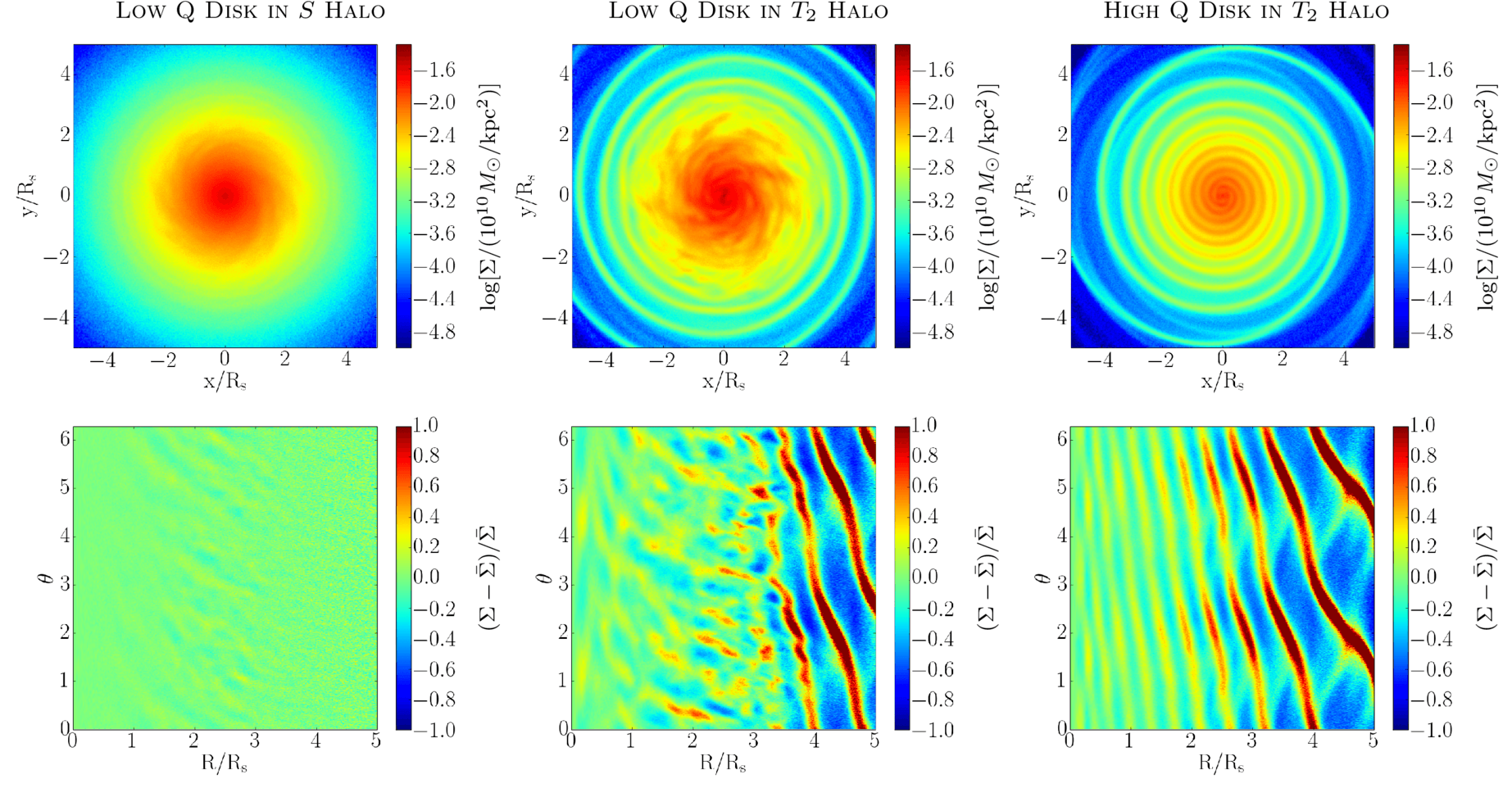}
  \caption{Surface density (top row) and residual density (bottom row) for
    discs in three different simulations at time $t=3\,\mathrm{Gyr}$. By
    comparing the left and the middle column we find that the strength of the
    transient spiral structures is stronger for the disc in a triaxial halo,
    although they are both formed due to swing amplification. As shown in the
    right column, transient multi-armed spiral structures do not form with a
    high-$Q$ disc.}
  \label{fig:lts}
\end{figure*}

\begin{figure*}
  \centering
\includegraphics[width=\linewidth]{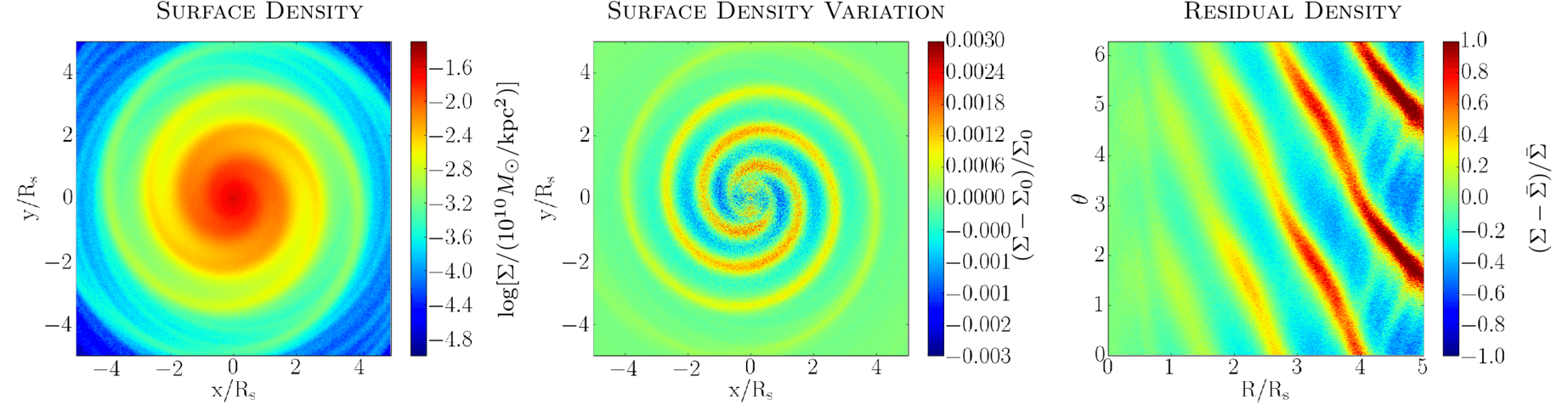}
  \caption{Simulation of low-$Q$ disc in $T_2$ halo without self-gravity at
    time $t=3\mathrm{Gyr}$. Without self-gravity, transient spiral
    structures no longer form,  while the grand-design spiral structures still
    persist.}
  \label{fig:ng}
\end{figure*}

To further explore the interaction between grand-design spiral structures and
swing amplification, we compare the discs formed in three different
simulations: one with a low-$Q$ disc in a $T_2$ halo, one with a
low-$Q$
disc in a spherical halo, and one with a high-$Q$ disc in a $T_2$ halo. The
surface density and residual density of the discs in these simulations at
$t=3\,\mathrm{Gyr}$ are shown in Fig.~\ref{fig:lts}. By comparing the
simulations with low-$Q$ and high-$Q$ disc in $T_2$ haloes, one can see
that the transient arms do not form in the simulation with the high-$Q$ disc,
where the swing amplification is very weak. This proves that the swing
amplification is the reason for the formation of the transient spiral
structures seen in the central regions of the low-$Q$ disc.

We additionally performed a simulation with self-gravity of the disc turned
off with  a low-$Q$ disc in a $T_2$ halo, whose result is shown in
Fig.~\ref{fig:ng}. No transient spiral structures form in this simulation,
although two-armed grand-design spiral structures still form. Without
self-gravity, there is no swing amplification process in the disc. This
further demonstrates that transient arms are caused by the swing
amplification, while two-armed grand-design spiral structures are not formed
due to the swing amplification.

\begin{figure*}
  \centering
\includegraphics[width=\linewidth]{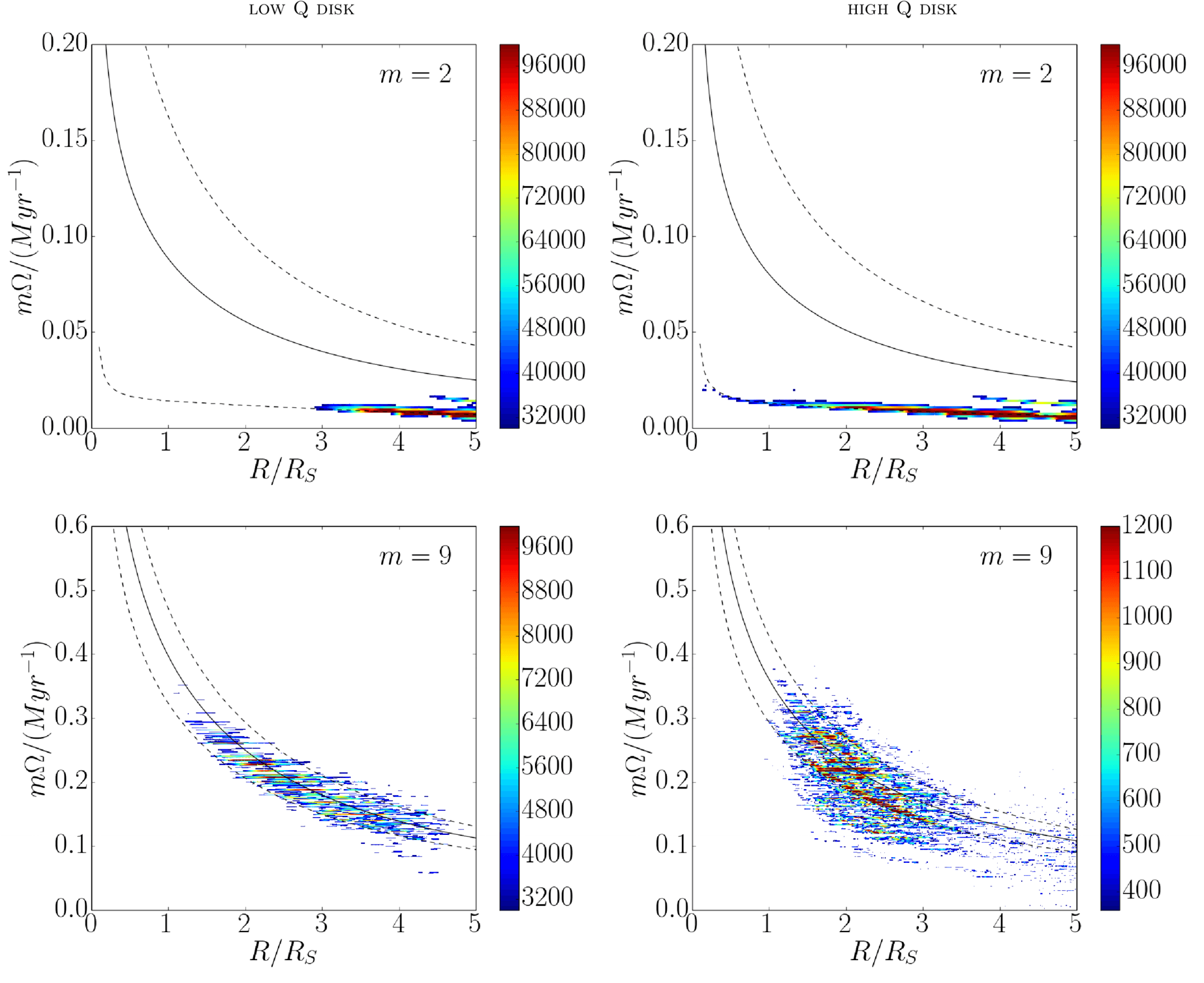}
  \caption{Power spectra of the low-$Q$ disc (left) and the high-$Q$ disc
    (right) for a time interval $2$--$7.5\,\mathrm{Gyr}$ in $T_2$ haloes for
    $m=2$ modes (top) and $m=9$ modes
    (bottom). In each panel, black solid curve indicates corotation line with
    pattern speed  $\Omega_{\rm P}=\Omega$, where $\Omega$ is the angular
    velocity of the stars, while black dashed curves are inner and outer
    Lindblad resonances with $\Omega_{\rm P}=\Omega\pm \kappa/2$, with $\kappa$
    being the epicyclic frequency. For $m=2$, the dominating feature
    corresponds to the grand-design spiral structures. For the low-$Q$ disc
    this structure only exists in the outer region, because the grand-design
    spiral structures are distorted by swing amplification. For the high-$Q$
    disc where swing-amplification is weak, the grand-design spiral modes
    extend down to the innermost regions. The grand-design spiral structures
    lie at the inner Lindblad resonance. The dominating features of $m=9$
    modes  are the horizontal structures crossing the corotation line, which
    are the typical structures of swing amplified modes. The strength of these
    modes in high-$Q$ disc is roughly an order of magnitude lower than in the
    low-$Q$ disc, as can be seen from the colour bar. Other high-$m$ modes
    are not shown,  as they are all qualitatively similar to the $m=9$ power
    spectra.}
  \label{fig:ps1}
\end{figure*}

We also compare the simulations with low-$Q$ discs in $S$ and $T_2$ haloes
as shown in Fig.~\ref{fig:lts}.  Transient spiral structures form in both
cases. However, the transient spiral structures formed in the $T_2$ halo are
significantly stronger than that formed in the $S$ halo. For the simulation
with the $T_2$ halo, two-armed spiral structures formed in the central region
of the disc at early times are gradually swing amplified and form the first
generation of transient spiral structures later, while for the simulation with
the $S$ halo, transient spiral structures form due to the swing amplification
of the Poisson noise in the initial conditions. With $10^8$ particles in the
disc, the Poisson noise is much weaker perturbation to the density field of
the disc than grand-design spiral structures. Therefore, transient spiral
structures formed in a $T_2$ halo are much stronger than that in an $S$ halo.

\subsection{Formation Mechanism of Two-Armed Spiral Structures}
\label{sec:formation-mechanism}
\begin{figure*}
    \centering
\includegraphics[width=.33\textwidth]{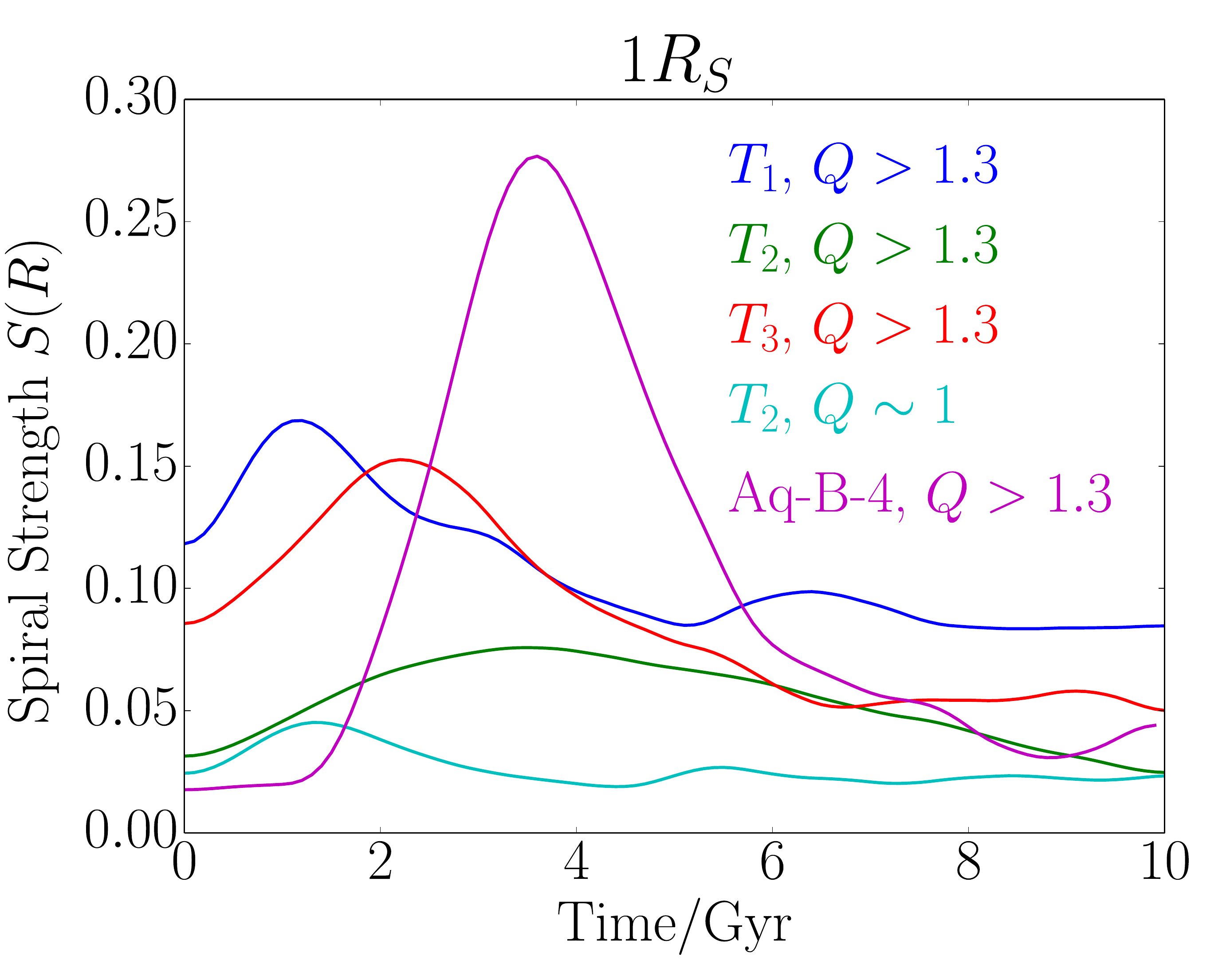}\includegraphics[width=.33\textwidth]{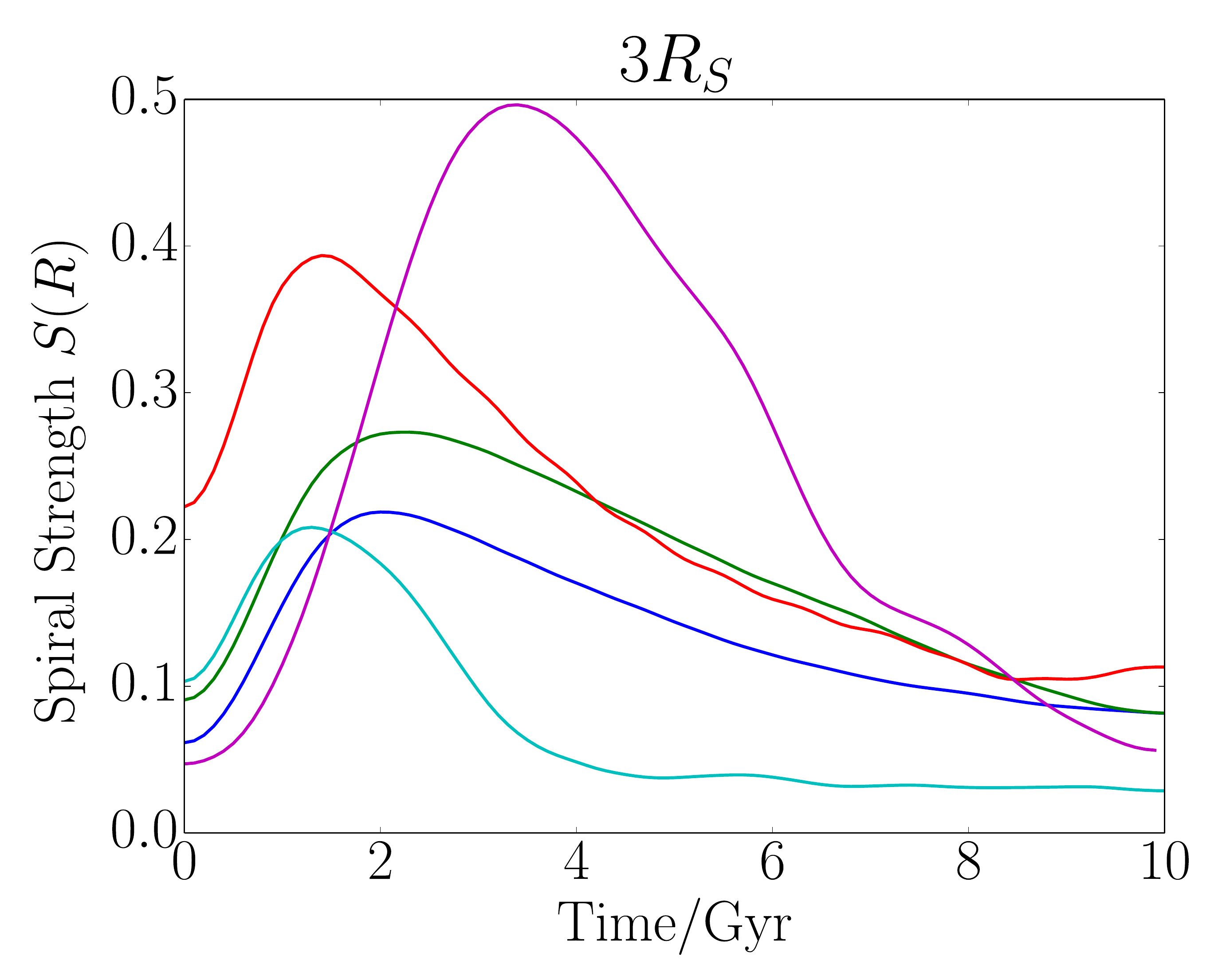}\includegraphics[width=.33\textwidth]{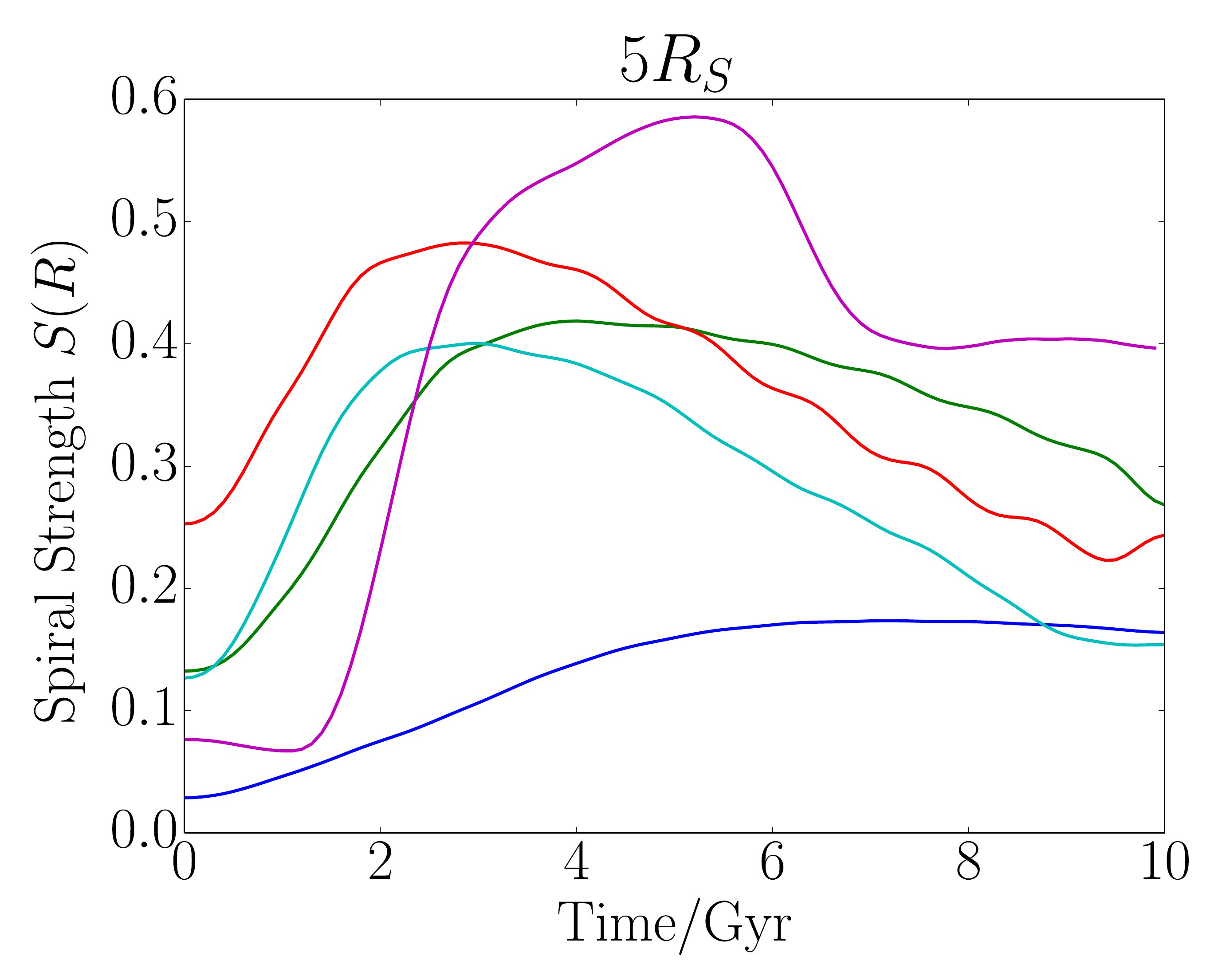}    
  \caption{The strength of the $m=2$ spiral structures as a function of time
    for radii $R=1R_\mathrm{S}, 3R_\mathrm{S}$ and $5R_\mathrm{S}$ in various
    simulations. For simulations with $T_1$, $T_2$, and $T_3$, the
      halo is static and triaxial, while for the simulation with the Aq-B-4
      halo (second model), the halo shape
      changes over time as in the right-hand panel of Fig. \ref{fig:realhalo}. For
      better comparison the time of the Aq-B-4 simulation has been shifted so that
    $t=0$ corresponds to $t=4\,\mathrm{Gyr}$ in Fig.~\ref{fig:realhalo}. Curves have been smoothed with a Hann function over a $\sim 2\,\mathrm{Gyr}$
    time span to show the general trends more clearly. For simulations with
    high-$Q$ discs, spiral strengths peak at $1$--$4\,\mathrm{Gyr}$ and then
    gradually decrease over several Gyr. In the $T_2$ halo with low-$Q$
    disc the strength of grand-design spirals is affected by transient spirals
    after $1-2\, \mathrm{Gyr}$. Simulation with the Aq-B-4 halo shows
      stronger spiral strength due to its higher triaxiality and fluctuating evolution. }  
  \label{fig:td}
\end{figure*}

To understand the formation mechanism of two-armed spiral structures, we study
the modes of spiral structures following \citet{sellwood2014}, by
measuring the power spectra of the discs in $T_2$ haloes.  The surface density
$\Sigma$ of the disc at time $t$, radius $R$ and azimuthal angle $\theta$ can
be expressed by the summation of a series of waves
\begin{equation}
  \label{eq:77}
  \Sigma(R,\theta,t)=\sum_{\Omega_f} \sum_mA(R,m,\Omega_f)\mathrm{e}^{\mathrm{i}(\Omega_f t+m\theta)}\,,
\end{equation}
where $m$ is the number of arms and $\Omega_f=m\Omega_\mathrm{P}$ is $m$ times the
pattern speed $\Omega_\mathrm{P}$. In other words, $\Sigma$ is the Fourier
  transform of the power $A$. Values of $A(R,m,\Omega_f)$ are complex
numbers, as they are a combination of $\cos$ and $\sin$ waves. We are more
interested in the absolute value $|A(R,m,\Omega_f)|$, which is the total
strength of the mode of $m$ and $\Omega_f$ at $R$. The power spectra are a good
tool to study the waves in the disc, as they show the dominant rotational modes
across the disc. In our simulations, since the surface density in the central
region is much higher than that in the outer region, small fluctuations in the centre
can have a higher power than the spiral structures located further out. In
order to  show the power spectra of the spirals clearly, we calculate power
spectra of the residual surface density instead of the surface density.

Power spectra of low-$Q$ and high-$Q$ discs in $T_2$ haloes for a time
interval from $2$ to $7.5\,\mathrm{Gyr}$ are shown in
Fig.~\ref{fig:ps1}.  For the power spectra of $m=2$ modes, the highest power
lies at the inner Lindblad resonance where the pattern speed is
$\Omega_\mathrm{P}=\Omega-\kappa/2$, with $\Omega$ being the rotation speed of
the stars and $\kappa$ being the epicyclic frequency. Since the high
  power values occur within a narrow region across the disc, there is only one
  dominating rotating mode in the disc, with rotating speed of
  $\Omega-\kappa/2$. The high-power region in 
the power spectra of the $m=2$ modes of the high-$Q$ disc spans throughout
the disc, but the high-power region for the low-$Q$ disc is located only at
the outer region of the disc. This is because the inner region of the
low-$Q$ disc is disrupted by the swing amplification, as previously shown in
Fig.~\ref{fig:lts}. We have also checked the $m=2$ power spectra of the
  simulation with self-gravity turned off. The high-power region lies at the
  inner Lindblad resonance similarly to the simulation of high-$Q$ disc with self-gravity, indicating that
  self-gravity plays a minor role here. For the power spectra of $m=9$ modes, the power strength
of the low-$Q$ disc is more than five times higher than that of the
high-$Q$ disc, because the latter does not form prominent transient
multi-armed spiral structures. Horizontal structures present in the power
spectra for  the $m=9$ modes of the low-$Q$ disc are shown to be the
characteristic features of the transient spiral structures due to the  swing
amplification by \citet{sellwood2014}.

The fact that the regions of the highest power of $m=2$ modes lie at the
inner Lindblad resonance with and without self-gravity indicates that the two-armed grand-design spiral
structures are indeed kinematic density waves, as proposed by
\citet{Lindblad1956}. For a large region in the disc, the value of
$\Omega-\kappa/2$ in our simulations can be regarded as roughly constant. Due
to the epicyclic motion of the stars, 
in a  frame rotating with this angular speed, stars have stationary elliptical
orbits. Therefore, in such a frame if the orbits of the stars are arranged to
be more crowded in some regions than the others initially, the crowded regions
will remain crowded. Seen from a rest frame, this corresponds to a pattern
moving with the pattern speed $\Omega_\mathrm{P}=\Omega-\kappa/2$. In our
simulations, such arrangement of the orbits is caused by sudden introduction
of triaxial haloes. 
Once formed, such patterns can survive in the disc for
several Gyr (see also Fig.~\ref{fig:efo} and discussion in Section
\ref{sec:time-depend-triax}). It is also worth noticing that the curve of $\Omega -
  \kappa/2$ is not perfectly flat, as it slightly decreases
  with radius. Because of this, the
  pattern speed of the spiral structures is slightly decreasing with radius as
  well, which explains the winding of the spirals at later times, as shown in
  e.g. Fig.~\ref{fig:t21}.

 The lifetimes of grand-design spirals are shown in Fig.~\ref{fig:td}. Here
we plot the strength of the $m=2$ modes as a function of time for radii
$R=1R_\mathrm{S}, 3R_\mathrm{S}$ and $5R_\mathrm{S}$. For simulations with
high-$Q$ discs, in general, spiral strengths peak at $1\sim 4\,\mathrm{Gyr}$ and
then gradually decline. Spiral strengths larger than $0.1$ can persist for up
to $5-10\,\mathrm{Gyr}$, depending on the radius probed. The peak values
correlate strongly with the torque strength shown in Fig.~\ref{fig:pp},
i.e. the peak value of the spiral strength is stronger when the torque
strength is stronger. For the simulation with a low-$Q$ disc and a $T_2$
halo, at the beginning, the growth of the spiral strength is similar to the
simulation with the same halo but with a high-$Q$ disc, but the strength of the
spirals in the low-$Q$ disc starts to decrease earlier, in $1$--$2\,\mathrm{Gyr}$. This is the time when the low-$Q$ disc starts to develop
transient spiral structures due to swing amplification, which favours 
spiral structures with higher $m$ modes. Simulation with the halo which
  traces the Aq-B-4 triaxiality with time (second model) generally shows a similar trend as the
  simulations with static haloes, though it has a higher peak strength. This is caused by two effects: (a) the ratio of the axes of the Aq-B-4
  halo is generally lower and (b) the ratio is fluctuating.

\section{Conclusions}
\label{sec:conclusions}
In this paper, we used very high resolution $N$-body simulations to investigate
the influence of triaxial haloes on to stellar discs, especially on the
formation of spiral structures. In our simulations, the haloes are implemented
as analytic potentials rather than dark matter particles to minimize possible
numerical artefacts caused by Poisson noise and to allow us to perform many very
high resolution simulations of the stellar discs, which are represented with
up to $10^8$ particles.  While high-$Q$ discs are stable against spiral
structures in spherical haloes, we found that two-armed grand-design spiral
structures form in such discs if they are abruptly embedded within triaxial haloes. These
spiral structures extend all the way to the edge of the disc. Their strength
dependence with radius is determined by the torque from the triaxial halo
experienced by the disc.  We further showed that these spiral structures have
the following features.
\begin{itemize}
\item They do not form when the halo turns from spherical to triaxial
  adiabatically. This indicates that the impulsive introduction of triaxial
  haloes leads to the grand-design two-armed spiral structures.  
\item Realistic fluctuations in halo triaxiality with time, as predicted by
    cosmological simulations, also lead to very prominent grand-design two-armed spirals,
    as we have explicitly demonstrated using the Aquarius simulations
    \citep{Springel2008, Vera-ciro2011}.
\item They form in discs that have $Q>1.3$, i.e. when the swing amplification
  process is weak in the disc. This demonstrates that swing amplification
  is not necessary for the formation for these spiral structures.
\item They form even if the self-gravity of the disc is turned off, which
  again excludes swing amplification as the essential formation mechanism.
\item Once formed, they can survive in spherical haloes. Triaxial haloes are
  therefore not necessary for maintaining the spiral structure.
\item Their power spectra peak at the inner Lindblad resonance that satisfies
  $\Omega_\mathrm{P}=\Omega-\kappa/2$, which is almost flat for a large region of the
  disc in our model, in agreement with the kinematic density wave theory
  proposed by \citet{Lindblad1956}. This offers a way to test the
    occurrence of this grand-design arm formation mechanism by comparing to the
  observed pattern speeds.
\end{itemize}

Furthermore, we showed that the swing amplification process and the time
  variation of the halo triaxiality can in fact interfere with each
other. When the 
time-scale of the growth of transient spiral structures due to the swing
amplification is very short, which in our case occurs when the number of the
star particles used in the simulation is low, only transient arms form, while
their distribution becomes more asymmetric in a triaxial halo. When the
process of swing amplification is slow enough, two-armed grand-design spiral
structures form in the first place. However, these spiral structures break up
and get swing amplified into transient arms at later times. Moreover, at a
given resolution spiral structures formed in this way grow faster than those
formed in  a spherical halo where the Poisson noise in the initial conditions
is the only source for the swing amplification.

The sudden introduction of triaxiality into an initially spherical halo in
equilibrium with the stellar disc is clearly a very idealized setup and is not
directly applicable to observed systems. However, we have shown that similar
spiral structures develop as long as the time-scale of the gravitational
potential perturbation is shorter than the orbital period of
stars. Self-consistent cosmological simulations indicate that both inner and outer halo
triaxiality can change rapidly and significantly on short time-scales as
massive satellite flybys, mergers, large-scale inflows and tumbling
triaxial haloes are known to alter gravitational potentials rapidly and to
introduce torques. The formation of
spiral structures in these cases shares a similar underlying mechanism
with our simplified simulations, as we have verified explicitly, where
the strength of arms formed depends on the 
gravitational torques, and where grand-design arms form. Furthermore, if
stellar discs are subject to swing amplification, additional rapid
perturbations in gravitational potentials and torques might be conducive to
the development of transient arms.

\section*{Acknowledgements}
We thank the anonymous referee for a very constructive report and Cathie Clarke,
Christophe Pichon, Jim Pringle, Daisuke Kawata, 
Victor Debattista and Giuseppe Lodato for their useful comments and advice. SH
is supported by the CSC Cambridge Scholarship, jointly funded by the China
Scholarship Council and by the Cambridge Overseas Trust. DS acknowledges
support by the STFC and ERC Starting Grant 638707 `Black holes and their host galaxies:
co-evolution across cosmic time'. This work was performed on DiRAC Darwin
Supercomputer hosted by the University of Cambridge High Performance Computing
Service (http://www.hpc.cam.ac.uk/), provided by Dell Inc. using Strategic
Research Infrastructure Funding from the Higher Education Funding Council for
England and funding from the Science and Technology Facilities Council; DiRAC
Complexity system, operated by the University of Leicester IT Services, which
forms part of the STFC DiRAC HPC Facility (www.dirac.ac.uk). This equipment is
funded by BIS National E-Infrastructure capital grant ST/K000373/1 and STFC
DiRAC Operations grant ST/K0003259/1; COSMA Data Centric system at Durham
  University, operated by the Institute for Computational Cosmology on behalf
  of the STFC DiRAC HPC Facility. This equipment was funded by a BIS National
  E-infrastructure capital grant ST/K00042X/1, STFC capital grant
  ST/K00087X/1, DiRAC Operations grant ST/K003267/1 and Durham University.
DiRAC is part of the National E-Infrastructure.

\bibliographystyle{mn2e}
\bibliography{paper}

\begin{thebibliography}{}

\bibitem[\protect\citeauthoryear{{Athanassoula}}{{Athanassoula}}{2012}]{Athanassoula2012}
{Athanassoula} E.,  2012, \mnras, 426, L46

\bibitem[\protect\citeauthoryear{Barnes \& Efstathiou}{Barnes \&
  Efstathiou}{1987}]{barnes1987}
Barnes J.,  Efstathiou G.,  1987, \apj, 319, 575

\bibitem[\protect\citeauthoryear{{Bertin} \& {Lin}}{{Bertin} \&
  {Lin}}{1996}]{Bertin1996}
{Bertin} G.,  {Lin} C.~C.,  1996, {Spiral Structure in Galaxies a Density Wave
  Theory}.
Cambridge, MA MIT Press

\bibitem[\protect\citeauthoryear{{Binney}}{{Binney}}{1978}]{Binney1978}
{Binney} J.,  1978, \mnras, 183, 779

\bibitem[\protect\citeauthoryear{{Binney}, {Jiang} \& {Dutta}}{{Binney}
  et~al.}{1998}]{Binney1998}
{Binney} J.,  {Jiang} I.-G.,    {Dutta} S.,  1998, \mnras, 297, 1237

\bibitem[\protect\citeauthoryear{Bowden, Evans \& Belokurov}{Bowden
  et~al.}{2013}]{bowden2013}
Bowden A.,  Evans N.,    Belokurov V.,  2013, \mnras, 435, 928

\bibitem[\protect\citeauthoryear{{Bryan}, {Kay}, {Duffy}, {Schaye}, {Dalla
  Vecchia} \& {Booth}}{{Bryan} et~al.}{2013}]{Bryan2013}
{Bryan} S.~E.,  {Kay} S.~T.,  {Duffy} A.~R.,  {Schaye} J.,  {Dalla Vecchia} C.,
     {Booth} C.~M.,  2013, \mnras, 429, 3316

\bibitem[\protect\citeauthoryear{{Debattista}, {Moore}, {Quinn}, {Kazantzidis},
  {Maas}, {Mayer}, {Read} \& {Stadel}}{{Debattista}
  et~al.}{2008}]{Debattista2008}
{Debattista} V.~P.,  {Moore} B.,  {Quinn} T.,  {Kazantzidis} S.,  {Maas} R.,
  {Mayer} L.,  {Read} J.,    {Stadel} J.,  2008, \apj, 681, 1076

\bibitem[\protect\citeauthoryear{{Debattista}, {Ro{\v s}kar}, {Valluri},
  {Quinn}, {Moore} \& {Wadsley}}{{Debattista} et~al.}{2013}]{Debattista2013}
{Debattista} V.~P.,  {Ro{\v s}kar} R.,  {Valluri} M.,  {Quinn} T.,  {Moore} B.,
     {Wadsley} J.,  2013, \mnras, 434, 2971

\bibitem[\protect\citeauthoryear{DeBuhr, Ma \& White}{DeBuhr
  et~al.}{2012}]{debuhr2012}
DeBuhr J.,  Ma C.-P.,    White S.~D.,  2012, \mnras, 426, 983

\bibitem[\protect\citeauthoryear{{Dobbs} \& {Baba}}{{Dobbs} \&
  {Baba}}{2014}]{Dobbs2014}
{Dobbs} C.,  {Baba} J.,  2014, \pasa, 31, 35

\bibitem[\protect\citeauthoryear{{Dobbs} \& {Bonnell}}{{Dobbs} \&
  {Bonnell}}{2006}]{Dobbs2006}
{Dobbs} C.~L.,  {Bonnell} I.~A.,  2006, \mnras, 367, 873

\bibitem[\protect\citeauthoryear{{Dobbs} \& {Bonnell}}{{Dobbs} \&
  {Bonnell}}{2007}]{Dobbs2007}
{Dobbs} C.~L.,  {Bonnell} I.~A.,  2007, \mnras, 376, 1747

\bibitem[\protect\citeauthoryear{{D'Onghia}, {Springel}, {Hernquist} \&
  {Keres}}{{D'Onghia} et~al.}{2010}]{DOnghia2010}
{D'Onghia} E.,  {Springel} V.,  {Hernquist} L.,    {Keres} D.,  2010, \apj,
  709, 1138

\bibitem[\protect\citeauthoryear{D'Onghia, Vogelsberger \& Hernquist}{D'Onghia
  et~al.}{2013}]{DOnghia2013}
D'Onghia E.,  Vogelsberger M.,    Hernquist L.,  2013, \apj, 766, 34

\bibitem[\protect\citeauthoryear{{Dubinski}}{{Dubinski}}{1994}]{Dubinski1994}
{Dubinski} J.,  1994, \apj, 431, 617

\bibitem[\protect\citeauthoryear{Dubinski \& Chakrabarty}{Dubinski \&
  Chakrabarty}{2009}]{dubinski2009}
Dubinski J.,  Chakrabarty D.,  2009, \apj, 703, 2068

\bibitem[\protect\citeauthoryear{{Fouvry} \& {Pichon}}{{Fouvry} \&
  {Pichon}}{2015}]{Fouvry2015}
{Fouvry} J.-B.,  {Pichon} C.,  2015, \mnras, 449, 1982

\bibitem[\protect\citeauthoryear{{Fouvry}, {Pichon}, {Magorrian} \&
  {Chavanis}}{{Fouvry} et~al.}{2015}]{fouvry2015d}
{Fouvry} J.~B.,  {Pichon} C.,  {Magorrian} J.,    {Chavanis} P.~H.,  2015,
  \aap, 584, A129

\bibitem[\protect\citeauthoryear{{Franx} \& {de Zeeuw}}{{Franx} \& {de
  Zeeuw}}{1992}]{Franx1992}
{Franx} M.,  {de Zeeuw} T.,  1992, \apjl, 392, L47

\bibitem[\protect\citeauthoryear{Frenk, White, Davis \& Efstathiou}{Frenk
  et~al.}{1988}]{frenk1988}
Frenk C.~S.,  White S.~D.,  Davis M.,    Efstathiou G.,  1988, \apj, 327, 507

\bibitem[\protect\citeauthoryear{{Fujii}, {Baba}, {Saitoh}, {Makino}, {Kokubo}
  \& {Wada}}{{Fujii} et~al.}{2011}]{Fujii2011}
{Fujii} M.~S.,  {Baba} J.,  {Saitoh} T.~R.,  {Makino} J.,  {Kokubo} E.,
  {Wada} K.,  2011, \apj, 730, 109

\bibitem[\protect\citeauthoryear{Gittins \& Clarke}{Gittins \&
  Clarke}{2004}]{gittins2004}
Gittins D.~M.,  Clarke C.,  2004, \mnras, 349, 909

\bibitem[\protect\citeauthoryear{{Grand}, {Kawata} \& {Cropper}}{{Grand}
  et~al.}{2012a}]{Grand2012A}
{Grand} R.~J.~J.,  {Kawata} D.,    {Cropper} M.,  2012a, \mnras, 426, 167

\bibitem[\protect\citeauthoryear{{Grand}, {Kawata} \& {Cropper}}{{Grand}
  et~al.}{2012b}]{Grand2012}
{Grand} R.~J.~J.,  {Kawata} D.,    {Cropper} M.,  2012b, \mnras, 421, 1529

\bibitem[\protect\citeauthoryear{{Hahn}, {Teyssier} \& {Carollo}}{{Hahn}
  et~al.}{2010}]{Hahn2010}
{Hahn} O.,  {Teyssier} R.,    {Carollo} C.~M.,  2010, \mnras, 405, 274

\bibitem[\protect\citeauthoryear{Hernquist}{Hernquist}{1990}]{hernquist1990}
Hernquist L.,  1990, \apj, 356, 359

\bibitem[\protect\citeauthoryear{Jing, Mo, B{\"o}rner \& Fang}{Jing
  et~al.}{1995}]{jing1995}
Jing Y.,  Mo H.,  B{\"o}rner G.,    Fang L.,  1995, \mnras, 276, 417

\bibitem[\protect\citeauthoryear{Jing \& Suto}{Jing \& Suto}{2002}]{jing2002}
Jing Y.,  Suto Y.,  2002, \apj, 574, 538

\bibitem[\protect\citeauthoryear{{Julian} \& {Toomre}}{{Julian} \&
  {Toomre}}{1966}]{Julian1966}
{Julian} W.~H.,  {Toomre} A.,  1966, \apj, 146, 810

\bibitem[\protect\citeauthoryear{Kalnajs}{Kalnajs}{1973}]{kalnajs1973}
Kalnajs A.,  1973, in Proceedings of the Astronomical Society of Australia
  Vol.~2, Spiral structure viewed as a density wave.
p.~174

\bibitem[\protect\citeauthoryear{{Khoperskov}, {Khoperskov}, {Zasov}, {Bizyaev}
  \& {Khrapov}}{{Khoperskov} et~al.}{2013}]{Khoperskov2013}
{Khoperskov} A.~V.,  {Khoperskov} S.~A.,  {Zasov} A.~V.,  {Bizyaev} D.~V.,
  {Khrapov} S.~S.,  2013, \mnras, 431, 1230

\bibitem[\protect\citeauthoryear{{Khoperskov} \& {Bertin}}{{Khoperskov} \&
  {Bertin}}{2015}]{Khoperskov2015}
{Khoperskov} S.~A.,  {Bertin} G.,  2015, \mnras, 451, 2889

\bibitem[\protect\citeauthoryear{Lin \& Shu}{Lin \& Shu}{1964}]{lin1964}
Lin C.,  Shu F.~H.,  1964, \apj, 140, 646

\bibitem[\protect\citeauthoryear{{Lindblad}}{{Lindblad}}{1956}]{Lindblad1956}
{Lindblad} B.,  1956, Stockholms Observatoriums Annaler, 19, 7

\bibitem[\protect\citeauthoryear{{Lindblad}}{{Lindblad}}{1963}]{Lindblad1963}
{Lindblad} B.,  1963, Stockholms Observatoriums Annaler, 22, 5

\bibitem[\protect\citeauthoryear{{Purcell}, {Bullock}, {Tollerud}, {Rocha} \&
  {Chakrabarti}}{{Purcell} et~al.}{2011}]{Purcell2011}
{Purcell} C.~W.,  {Bullock} J.~S.,  {Tollerud} E.~J.,  {Rocha} M.,
  {Chakrabarti} S.,  2011, \nat, 477, 301

\bibitem[\protect\citeauthoryear{{Romeo}}{{Romeo}}{1992}]{romeo1992}
{Romeo} A.~B.,  1992, \mnras, 256, 307

\bibitem[\protect\citeauthoryear{{Salo}, {Laurikainen}, {Buta} \&
  {Knapen}}{{Salo} et~al.}{2010}]{Salo2010}
{Salo} H.,  {Laurikainen} E.,  {Buta} R.,    {Knapen} J.~H.,  2010, \apjl, 715,
  L56

\bibitem[\protect\citeauthoryear{{Sellwood}}{{Sellwood}}{2012}]{Sellwood2012}
{Sellwood} J.~A.,  2012, \apj, 751, 44

\bibitem[\protect\citeauthoryear{{Sellwood} \& {Carlberg}}{{Sellwood} \&
  {Carlberg}}{2014}]{sellwood2014}
{Sellwood} J.~A.,  {Carlberg} R.~G.,  2014, \apj, 785, 137

\bibitem[\protect\citeauthoryear{Springel}{Springel}{2005}]{springel2005cosmological}
Springel V.,  2005, \mnras, 364, 1105

\bibitem[\protect\citeauthoryear{Springel, Di~Matteo \& Hernquist}{Springel
  et~al.}{2005}]{springel2005}
Springel V.,  Di~Matteo T.,    Hernquist L.,  2005, \mnras, 361, 776

\bibitem[\protect\citeauthoryear{Springel, Wang, Vogelsberger, Ludlow, Jenkins,
  Helmi, Navarro, Frenk \& White}{Springel et~al.}{2008}]{Springel2008}
Springel V.,  Wang J.,  Vogelsberger M.,  Ludlow A.,  Jenkins A.,  Helmi A.,
  Navarro J.~F.,  Frenk C.~S.,    White S.~D.,  2008, \mnras, 391, 1685

\bibitem[\protect\citeauthoryear{Thomas, Colberg, Couchman, Efstathiou, Frenk,
  Jenkins, Nelson, Hutchings, Peacock, Pearce et~al.,}{Thomas
  et~al.}{1998}]{thomas1998}
Thomas P.~A.,  Colberg J.~M.,  Couchman H.~M.,  Efstathiou G.~P.,  Frenk C.~S.,
   Jenkins A.~R.,  Nelson A.~H.,  Hutchings R.~M.,  Peacock J.~A.,  Pearce
  F.~R.,    et~al., 1998, \mnras, 296, 1061

\bibitem[\protect\citeauthoryear{Toomre}{Toomre}{1964}]{toomre1964}
Toomre 1964, \apj, 139, 1217

\bibitem[\protect\citeauthoryear{Toomre}{Toomre}{1981}]{toomre1981}
Toomre 1981, in Structure and Evolution of Normal Galaxies Vol.~1, What
  amplifies the spirals.
pp 111--136

\bibitem[\protect\citeauthoryear{{van de Voort}, {Davis}, {Kere{\v s}},
  {Quataert}, {Faucher-Gigu{\`e}re} \& {Hopkins}}{{van de Voort}
  et~al.}{2015}]{vandeVoort2015}
{van de Voort} F.,  {Davis} T.~A.,  {Kere{\v s}} D.,  {Quataert} E.,
  {Faucher-Gigu{\`e}re} C.-A.,    {Hopkins} P.~F.,  2015, \mnras, 451, 3269

\bibitem[\protect\citeauthoryear{{Vandervoort}}{{Vandervoort}}{1970}]{vandervoort1970}
{Vandervoort} P.~O.,  1970, \apj, 161, 87

\bibitem[\protect\citeauthoryear{{Vera-Ciro}, {Sales}, {Helmi}, {Frenk},
  {Navarro}, {Springel}, {Vogelsberger} \& {White}}{{Vera-Ciro}
  et~al.}{2011}]{Vera-ciro2011}
{Vera-Ciro} C.~A.,  {Sales} L.~V.,  {Helmi} A.,  {Frenk} C.~S.,  {Navarro}
  J.~F.,  {Springel} V.,  {Vogelsberger} M.,    {White} S.~D.~M.,  2011,
  \mnras, 416, 1377

\bibitem[\protect\citeauthoryear{Warren, Quinn, Salmon \& Zurek}{Warren
  et~al.}{1992}]{warren1992}
Warren M.~S.,  Quinn P.~J.,  Salmon J.~K.,    Zurek W.~H.,  1992, \apj, 399,
  405

\bibitem[\protect\citeauthoryear{Yoshida, Springel, White \& Tormen}{Yoshida
  et~al.}{2000}]{yoshida2000}
Yoshida N.,  Springel V.,  White S.~D.,    Tormen G.,  2000, \apj Letters, 544,
  L87

\bibitem[\protect\citeauthoryear{Zemp, Gnedin, Gnedin \& Kravtsov}{Zemp
  et~al.}{2012}]{zemp2012}
Zemp M.,  Gnedin O.~Y.,  Gnedin N.~Y.,    Kravtsov A.~V.,  2012, \apj, 748, 54

\bibitem[\protect\citeauthoryear{{Zhu}, {Marinacci}, {Maji}, {Li}, {Springel}
  \& {Hernquist}}{{Zhu} et~al.}{2016}]{Zhu2015}
{Zhu} Q.,  {Marinacci} F.,  {Maji} M.,  {Li} Y.,  {Springel} V.,    {Hernquist}
  L.,  2016, \mnras, 458, 1559

\end{thebibliography}

\appendix

\section{Disc Stability in Spherical Halo}
\label{app:dssh}

\begin{figure*}
\includegraphics[width=\linewidth]{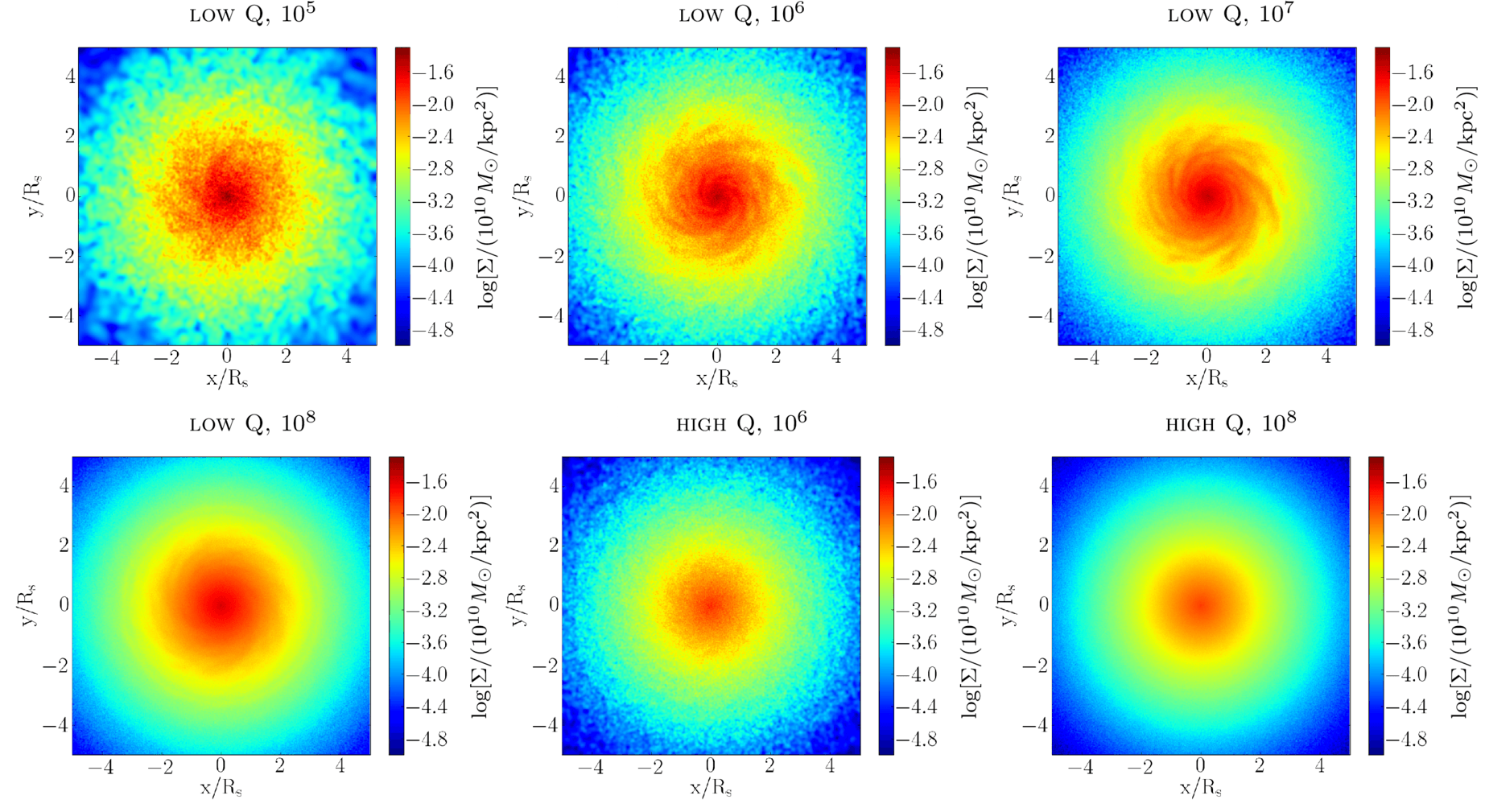}
    \caption{Surface density of the low-$Q$ and high-$Q$ stellar disc  at time
      $t = 2.5\mathrm{Gyr}$. For low-$Q$ discs, four simulations with increasing number of
      particles, ranging from $10^5$  to $10^8$, are
      shown. Strong spiral structures develop when the particle number is
      lower than $10^8$, while for the disc with $10^8$ particles, very weak
      spiral structures can also be seen. For $10^5$ particles, the spiral
      structures initially present fade away before $t = 2.5\mathrm{Gyr}$.
      For high-$Q$ discs, simulations with $10^6$ and $10^8$ star
        particles are shown. No prominent spiral structures can be found in such
      simulations.}
    \label{fig:1}
  \end{figure*}

\citet{Fujii2011}, \citet{Sellwood2012}, \citet{DOnghia2013} and several other
works have found that  the  Poisson noise in the disc due to a low number of
star particles can be greatly swing amplified causing transient spiral
structures to form. As stated in Section~\ref{sec:finite-resol-effects}, we
performed several simulations to study this effect. Two competing
  factors, the initial Poisson noise level set by the number of particles
  $N$ and the gravitational stability of the disc quantified by Toomre's $Q$
  parameter, are considered. Our simulations include
two runs with a high-$Q$ disc containing $10^6$ and $10^8$ star particles and four
runs with low-$Q$ discs containing $10^5$, $10^6$, $10^7$ and $10^8$
star particles. Surface densities of discs at $2.5\,\mathrm{Gyr}$ are shown in
Fig.~\ref{fig:1}. For high-$Q$ discs, no spiral structures form. For discs with low-$Q$, strong spiral structures form in simulation with $10^6$
and $10^7$ star particles, while very weak spiral structures are present in
simulation with $10^8$ star particles. For the simulation with low-$Q$ with $10^5$ star
particles, the spiral structures can barely be seen due to two reasons: (a) as
shown in Fig.~\ref{fig:svt}, the strength of spiral structures in this
simulation decreases after several hundred million years, (b) number of
particles is too low to show weak spiral structures.

For simulations with low-$Q$ discs with $10^6$--$10^8$ star particles, the spiral structures
are most prominent in the region $R_\mathrm{S}<r<3R_\mathrm{S}$. We checked
the evolution of such spiral arms and found that each single arm is not
long-lived. The arms break up quickly and new arms emerge from the fragments
of the old arms. The stars in the disc are rotating anticlockwise;
therefore, the spiral structures are trailing, which is in the agreement with
the prediction of the swing amplification mechanism. 

\begin{figure*}
  \begin{minipage}{\linewidth}
    \centering
    \includegraphics[width=.4\textwidth]{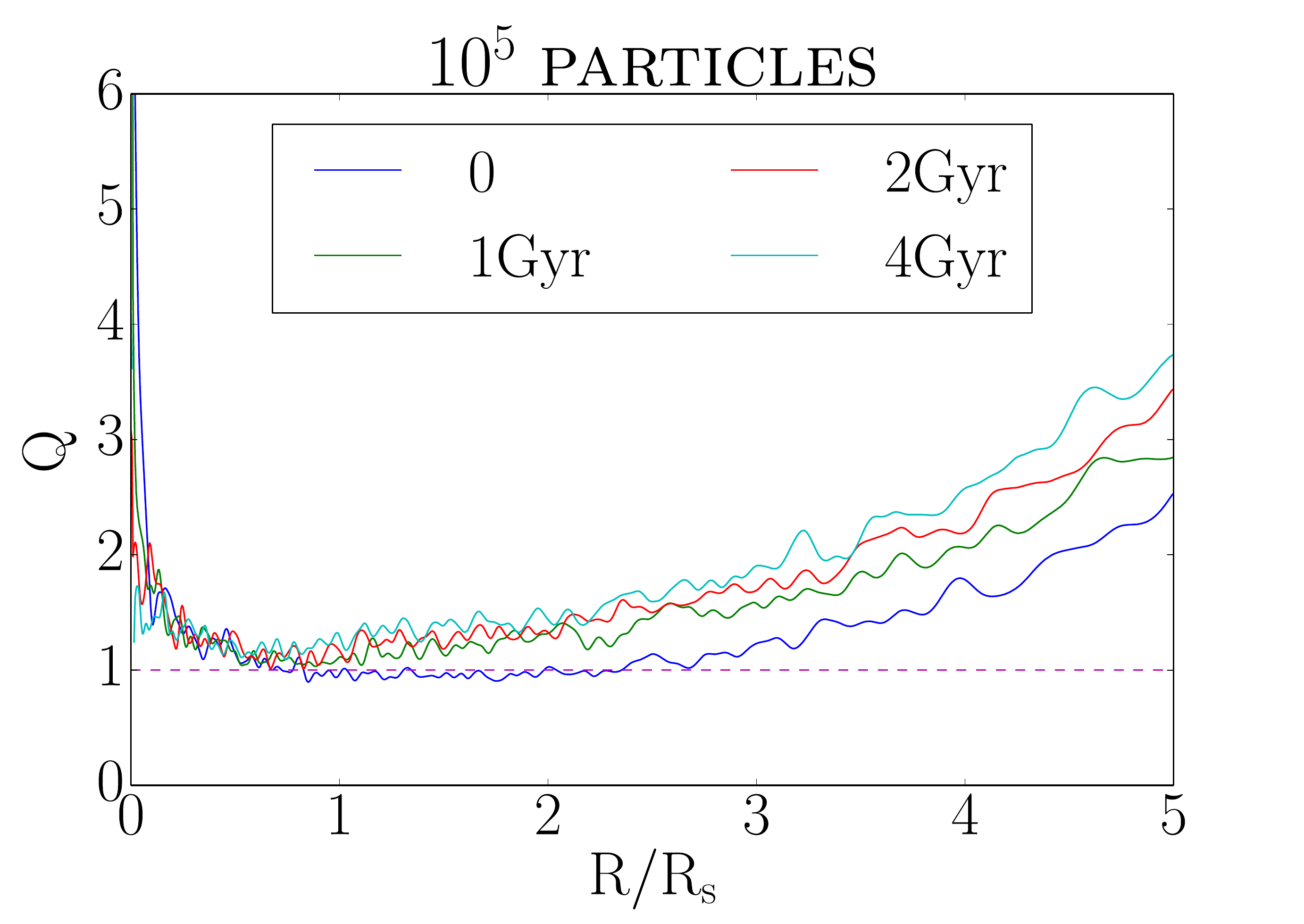} \includegraphics[width=.4\textwidth]{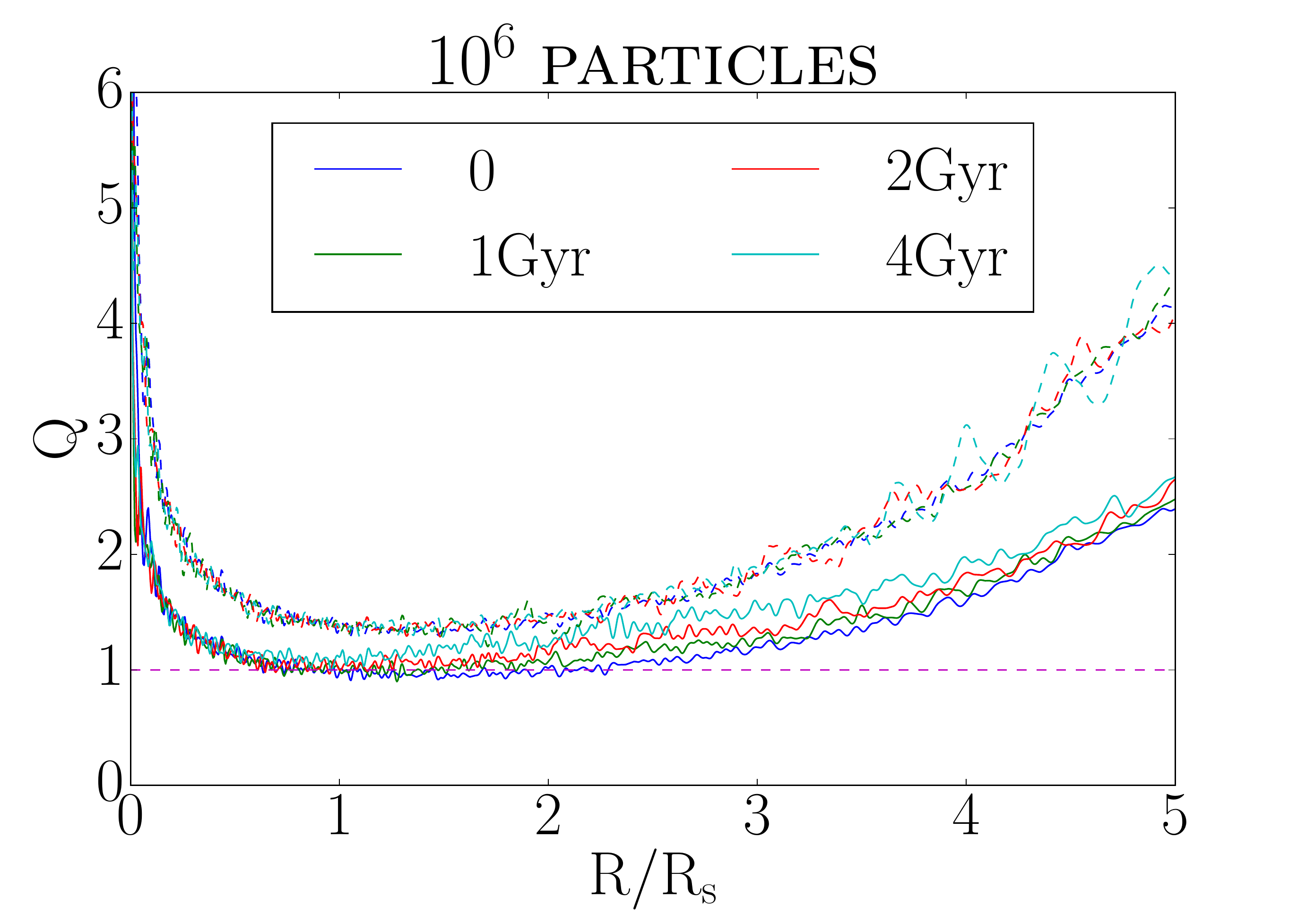}

    \includegraphics[width=.4\textwidth]{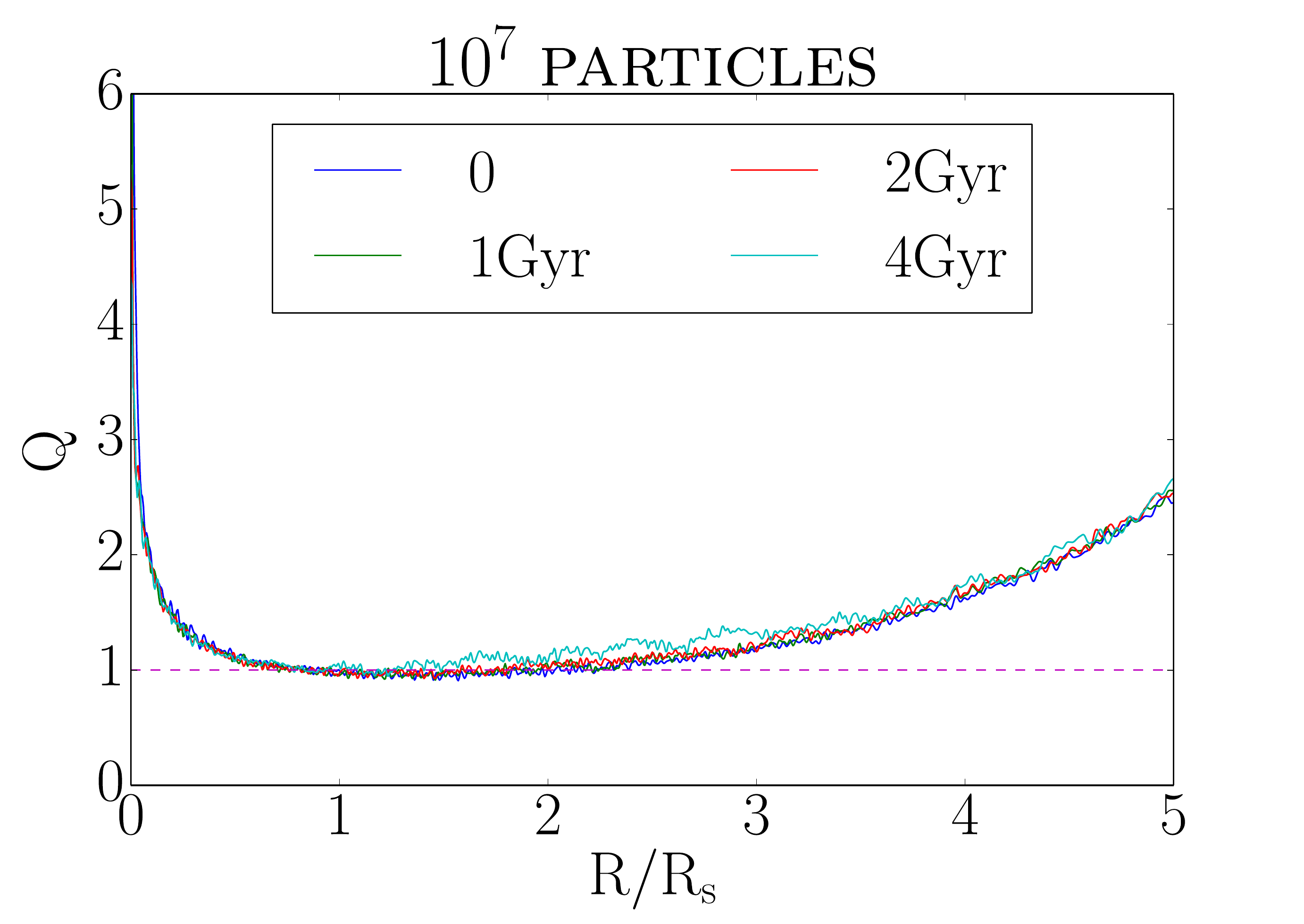} \includegraphics[width=.4\textwidth]{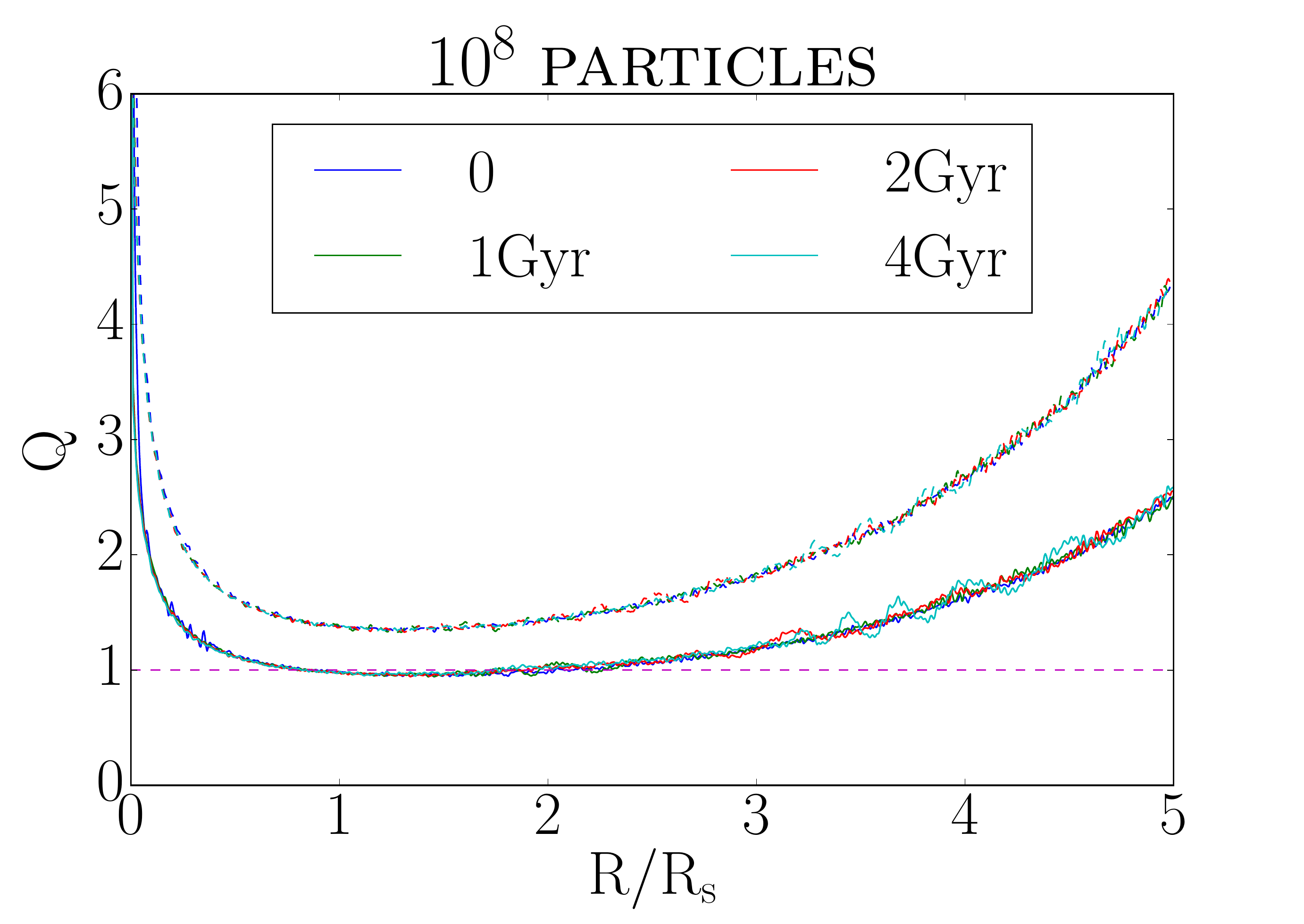}
  \caption{Toomre's $Q$ parameter as a function of radius. Plots for four
    identical low-$Q$ discs simulated with  different number of particles
    (from $10^5$ to $10^8$) are shown with solid lines, while high-$Q$
      discs with $10^6$ and $10^8$ particles are shown with dashed lines
    in the right-hand panels. Lines
    of different colour stand for different times, as annotated in legends. In
    particular, blue lines in each plot denote the $Q$--$r$ relation at the
    initial time.  The swing amplification is strong if $Q$ is close
    to 1. As spiral structures develop, the `temperature' of the disc raises
    for low particle numbers, i.e. $10^5$ and $10^6$, resulting in an increase
    of $Q$  at 
    later times. This increase of $Q$ coincides with the decrease of spiral
    strength of corresponding models shown in Fig.~\ref{fig:svt}.
    For high-$Q$ discs, no prominent spiral structures develop, leading to
    hardly any changes in $Q$ profile.}
  \label{fig:2}
  \end{minipage}
\end{figure*}

The average $Q$ value as a function of radius is shown in
Fig.~\ref{fig:2}. Blue lines denote the $Q$ value in the initial
conditions. $Q$ parameter for simulations with low-$Q$ discs is shown with
solid curves. For low-$Q$ discs, in the region of
$R_\mathrm{S}<r<3R_\mathrm{S}$, the $Q$ parameter is either less than $1$ or
slightly over $1$. Due to the fact that discs are not razor-thin, they
  are stable to axisymmetric perturbations even when $Q$ is slightly lower than 1, but the swing
  amplification is still strong. This explains why transient spiral structures are strongest
in this region. For simulation with a lower number of particles, $Q$
value grows faster
over time. At later times, $Q$ is so high that swing amplification is no
longer strong, which leads to the decreasing of spiral strength.  In 
simulations with discs with $Q > 1.3$, $Q$ parameter is higher than $1.3$ all
through the disc, as shown in the right-hand panels of Fig.~\ref{fig:2}
with dashed curves. Due to a high $Q$, spiral structures do not grow
prominently in the disc.

In conclusion, the growth of self-induced spirals depends on two
  factors: the initial Poisson noise level and the gravitational stability of
  the disc. The initial Poisson noise level sets the initial strength of
  perturbations that are amplified later. The higher number of particles, the longer
time it takes to grow perturbations to a prominent level. The stability of the
disc decides the actual growth rate of the perturbation. For a highly stable
disc, it may take very long time to grow perturbations in a self-induced manner.

\section{Discs in Misaligned Triaxial Haloes}
\label{sec:discs-misal-triax}
\begin{figure*}
  \centering
  \includegraphics[width=\linewidth]{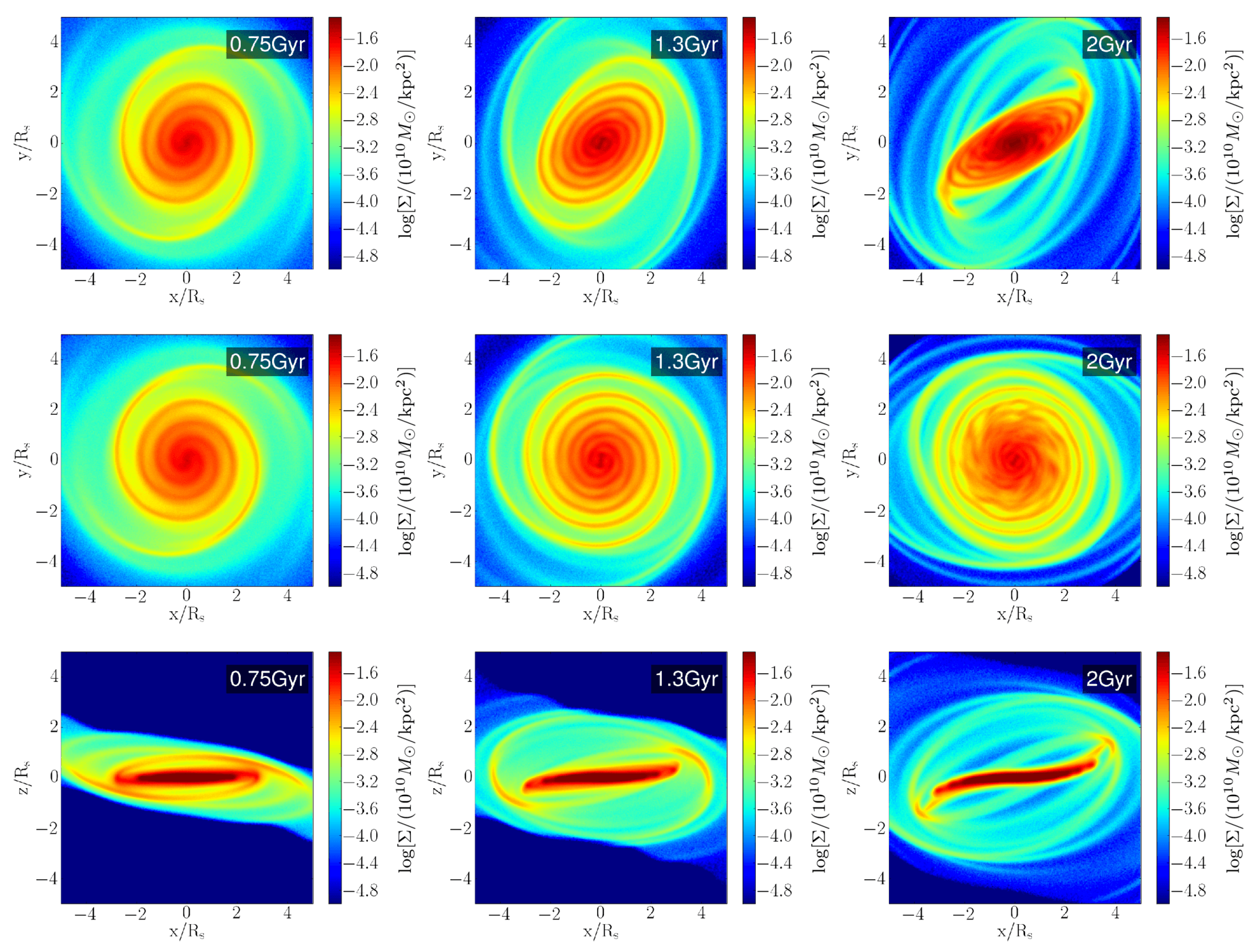}
  \caption{Evolution of a disc in a misaligned triaxial halo. Top:
    projection of density on the initial plane. Middle: projection of
    density on the disc plane, whose normal direction is defined by
    the total angular momentum. Bottom: edge-on column density of the
    disc. Grand-design two-armed spiral structures still form in
    this case, while the orientation of the disc is changing as well. It
    can be also seen from the edge-on plot that warps develop in the
    disc at later times.}
  \label{fig:to}
\end{figure*}

\begin{figure*}
\centering
\includegraphics[width=.97\linewidth]{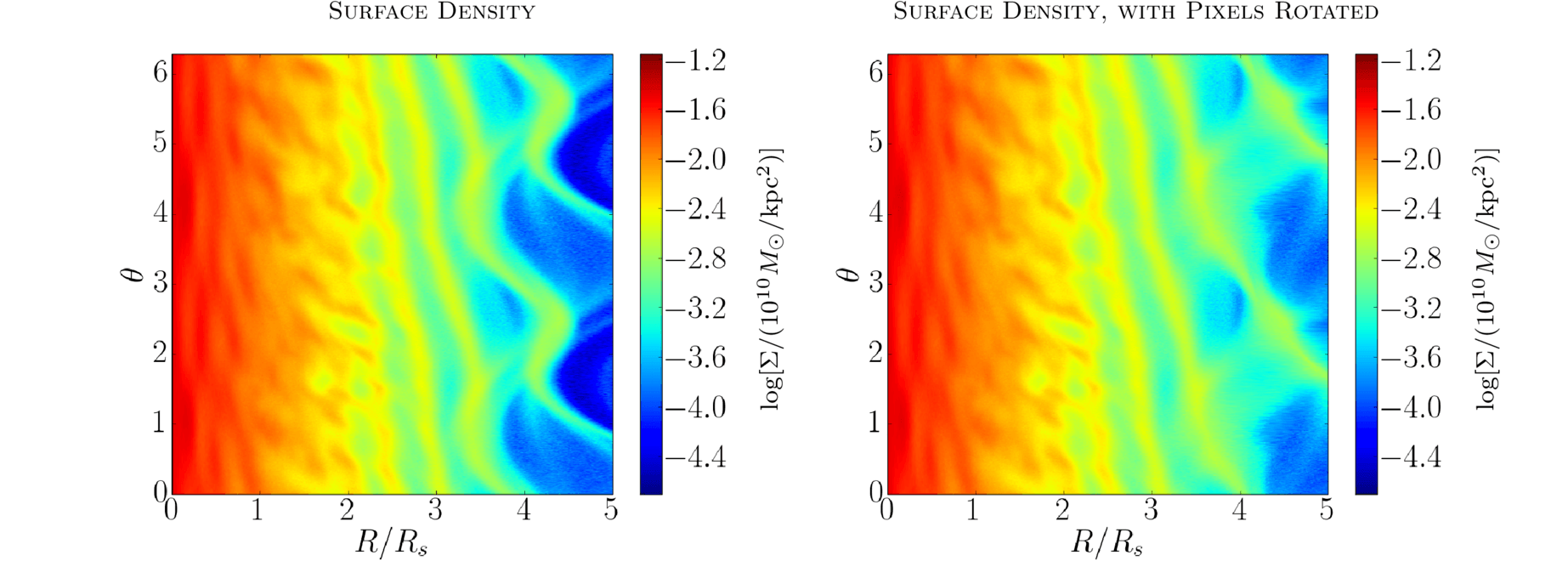}
  \caption{Surface density of a disc in a misaligned triaxial halo at
    $2\,\mathrm{Gyr}$ in polar 
    coordinates. The outer part of the disc is not in the
    plane defined by the angular momentum of the disc. The spiral arms are
    distorted if the density is directly projected on to the disc plane
    (left). However, if we rotate every pixel on to the disc plane (right), the
    morphology of the spiral structure becomes similar to the simulations with
    discs within the $x$--$y$ plane of the haloes.}
  \label{fig:sff}
\end{figure*}

We now focus on to the influence of triaxial haloes if the disc does not lie in
the $x$--$y$ plane of the halo.  Even though for isolated systems it has been
shown that the inner part of the halo realigns on a dynamical time with the
disc \citep{Binney1998}, in cosmological simulations large misalignments are
found \citep[e.g.][]{Hahn2010,Debattista2013, vandeVoort2015}, indicating that
discs in misaligned haloes need to be studied.
 The top row of Fig.~\ref{fig:to} shows the
projection of the disc on to its initial plane in the simulation with a
$T_\mathrm{O2}$ halo, i.e. a halo that is more triaxial outside with the disc
plane that initially has a 45$^\circ$ angle with respect to the $x$--$y$ plane of
the halo, at three different times, $t=0.7, 1.3$
and $2\,\mathrm{Gyr}$. At $0.7\,\mathrm{Gyr}$, the overall shape of the
projection starts to be compressed.  This  compression becomes stronger over
time. At around $t=2\,\mathrm{Gyr}$, the projection of the disc is strongly
compressed in one direction. This compression of the projection on to a fixed
plane indicates that the inclination of the disc is changing. In a triaxial
halo, the gravity force does not always point directly to the centre. It does
not even lie in the disc plane if the disc plane does not coincide with the $x$--$y$,
$x$--$z$ or $y$--$z$ plane of the halo. The perpendicular component of the gravity
force results in a non-zero torque that is not in the direction of angular
momentum, which leads to the change of the disc's inclination.

We can infer the normal direction of the disc with the direction of the
angular momentum vector, as long as the stars stay roughly in a plane. To
achieve, this we calculate the total angular momentum
\begin{equation}
\bm{L}=\sum_i m \bm{r}_i\times \bm{v}_i
\end{equation}
at each timestep, where $m$ is the mass of a single particle, $\bm{r}_i$
and $\bm{v}_i$ are the position and velocity vector of the $i$-th star
particle.

The middle row of Fig.~\ref{fig:to} shows the projection of the disc onto
the plane that is perpendicular to the total angular  momentum vector. We can
see that the morphology of the spiral structures is  similar to the disc that
is not misaligned with respect to the $x$--$y$ plane of the halo, for example
similar to the disc shown in  Fig.~\ref{fig:lt}. However, in the outer parts
of the disc the shape of the spiral arms is distorted.

We can understand the distortion of the disc by looking at it edge-on. As
shown in the bottom row of Fig.~\ref{fig:to}, the disc develops an integral
shape warp at later times, indicating that the disc no longer stays in a
plane. Also more and more mass spreads along the $z$ direction. It is also
worth noticing that though the warping of the disc shows a trend for the disc
to become aligned with the major axis of the halo, which is $45^\circ$ from
the initial normal direction of the disc, in fact most of the mass of the disc
stays in roughly the same initial plane at time $t=2\,\mathrm{Gyr}$.

The warp of the disc before $0.75\,\mathrm{Gyr}$ is very weak. Therefore we
can compare the strength of the spiral structures at early times,
e.g. $0.5\,\mathrm{Gyr}$, in this simulation with the simulations where the
discs lie in the $x$--$y$ plane of the triaxial halo, as shown in the bottom row
of Fig.~\ref{fig:3} and the right-hand panel of Fig.~\ref{fig:pp}. The strength
of the spiral structures in this simulation also matches with the
gravitational torque caused by the triaxial halo. In fact, the strength of the
torque in this simulation lies between the simulations with $T_2$ and $T_3$
haloes, as well as the corresponding strength  of the spiral structures.

At later times, the outer part of the disc no longer stays in the disc
plane. As shown in the middle row of Fig.~\ref{fig:to}, the outer part of
the spiral structures is distorted. We plot the density projection of the disc
in polar coordinates, as illustrated in the left-hand panel of
Fig.~\ref{fig:sff}. For $R/R_\mathrm{S}>3.5$, radius of the spiral
structures no longer increases strictly as the angle goes
anticlockwise. To investigate if this is due to the warp of the disc, we
rotate every pixel in this plot on to the disc plane again while keeping the
azimuthal coordinate $\theta$ unchanged, so that the spiral structure all
through the disc can be compared on the same plane. This is done by changing
the radius $R$ of each pixel to $\sqrt{R^2+\overline{z}^2}$, where 
\begin{equation}
  \label{eq:2}
  \overline{z}=\frac{1}{\Sigma(R,\theta)}\int_{-\infty}^\infty \rho(R,\theta,z)z\mathrm{d}z\,,
\end{equation}
is the mean height of the mass at $(R,\theta)$. The result in shown in the
right-hand panel of Fig.~\ref{fig:sff}. The radius of the spiral structures
increases strictly anticlockwise, as is the case for the spiral structures
in all other simulations  with discs in the $x$--$y$ plane of the halo. This
indicates that the distortion of the outer part of the spiral structures is
simply due to the projection.

In conclusion, when the disc is misaligned with the halo, the spiral
structures still grow in response to the torque. Their strength corresponds to
the strength of the torque, similar to the simulations with discs lying in the
halo plane. Even though warps develop at later times, they interfere very
little with the spiral structures. The morphology of the spiral structures is
unaffected as long as the radii of the spiral structures are calculated with
the height from the disc taken into account.

\bsp	
\label{lastpage}
\end{document}